\documentclass[twocolumn]{aastex631}
\usepackage[utf8]{inputenc}
\usepackage{float}
\usepackage{amsmath, bm, bbm}
\usepackage{amssymb}
\usepackage{physics}
\usepackage{tikz}
\usepackage{enumitem}
\DeclareRobustCommand{\bbone}{\text{\usefont{U}{bbold}{m}{n}1}}
\DeclareMathOperator{\EX}{\mathbb{E}}

\begin{document}

\title{Pixelated Reconstruction of Foreground Density and Background Surface Brightness in Gravitational Lensing Systems using Recurrent Inference Machines}

\author{\href{https://orcid.org/0000-0001-8806-7936}{\includegraphics[width=8pt]{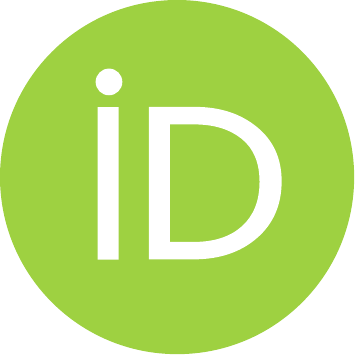}} Alexandre Adam }
\affiliation{Department of Physics, Université de Montréal, Montréal, Canada}
\affiliation{Mila - Quebec Artificial Intelligence Institute, Montréal, Canada}
\affiliation{Ciela - Montreal Institute for Astrophysical Data Analysis and Machine Learning, Montréal, Canada}

\author{\href{https://orcid.org/0000-0003-3544-3939}{\includegraphics[width=8pt]{orcid.pdf}} Laurence Perreault-Levasseur }
\affiliation{Department of Physics, Université de Montréal, Montréal, Canada}
\affiliation{Mila - Quebec Artificial Intelligence Institute, Montréal, Canada}
\affiliation{Ciela - Montreal Institute for Astrophysical Data Analysis and Machine Learning, Montréal, Canada}
\affiliation{Center for Computational Astrophysics, Flatiron Institute, 162 5th Avenue, 10010, New York, NY, USA}

\author{ \href{https://orcid.org/0000-0002-8669-5733}{\includegraphics[width=8pt]{orcid.pdf}} Yashar Hezaveh}
\affiliation{Department of Physics, Université de Montréal, Montréal, Canada}
\affiliation{Mila - Quebec Artificial Intelligence Institute, Montréal, Canada}
\affiliation{Ciela - Montreal Institute for Astrophysical Data Analysis and Machine Learning, Montréal, Canada}
\affiliation{Center for Computational Astrophysics, Flatiron Institute, 162 5th Avenue, 10010, New York, NY, USA}

\author{Max Welling}
\affiliation{Microsoft Research AI4Science}

\begin{abstract}
Modeling strong gravitational lenses in order to 
quantify the distortions in the images of background sources and 
to reconstruct the mass density in the foreground lenses has been a difficult computational challenge. 
As the quality of gravitational lens images increases, the task of fully exploiting the information they contain 
becomes computationally and algorithmically more difficult. 
In this work, we use a neural network based on the Recurrent Inference Machine (RIM) to simultaneously reconstruct an undistorted image of the background source and the lens mass density distribution as pixelated maps. 
The method iteratively reconstructs the model parameters (the image of the source and a pixelated density map) by learning 
the process of optimizing the likelihood given the data using the physical model (a ray-tracing simulation), regularized
by a prior implicitly learned by the neural network through its training data. When compared to more traditional parametric models, 
the proposed method is significantly more expressive and can reconstruct complex mass distributions, 
which we demonstrate by using realistic lensing galaxies taken from the IllustrisTNG cosmological hydrodynamic simulation. 
\end{abstract}



\section{Introduction}

Strong gravitational lensing is a natural phenomenon through which multiple, distorted images of luminous background sources are formed by the gravity of  massive foreground objects along the line of sight 
\citep[e.g.,][]{Viera2013,Marrone2018,Rizzo2020,Sun2021}. 
These distortions are tracers of the distribution of mass in foreground structures, irrespective of their light emission properties. 
As such, this phenomenon offers a powerful probe of the distribution of 
dark matter \citep[e.g.,][]{Dala2002,Treu2004,Hezaveh2016,Gilman2020,Gilman2021}.

Lens modeling is the process through which the parameters describing both the mass distribution in the 
foreground lens and the undistorted image of the background source are inferred.
This has traditionally been done through explicit likelihood-based modeling methods, a time- and resource-consuming procedure. 
A common practice to model strong lenses is to model the light profile of the background source with a \citet{Sersic1963} profile and the density of the foreground lens with a power law function, $\rho \propto r^{-\gamma}$. 
These simple profiles allow for the exploration of their low-dimensional parameter space with non-linear samplers such as Markov Chain Monte Carlo (MCMC) methods \citep[e.g.,][]{Koopmans2006,Barnabe2009,Auger2010} and generally provide a good fit to low-resolution data. 
However, as high-resolution and high signal-to-noise ratio (SNR) images become available, lensing analysis with simple models requires the introduction of additional parameters representing the true complexity of the mass distribution in lensing galaxies and the complexity of surface brightness in the background sources \citep[e.g.,][]{Sluse2017,Wong2017,Birrer2019,Rusu2019, Rusu2017,Li2021}. 
This approach becomes intractable as the complexity of the mass distribution and the quality of images increases \citep[e.g.,][]{Schmidt2022}. For example,
no simple parametric model of the \textit{Hubble Space Telescope} (\textit{HST}) images of the Cosmic Horseshoe (J1148+1930) --- initially discovered by \citet{Belokurov2007} --- has been able to model the fine features of the extended arc \citep[e.g.,][]{Bellagamba2016,Cheng2019,Schuldt2019}.

Free-form methods attempt to relax the assumptions about the smoothness and symmetries of these parametric profiles
using more expressive families like regular (or adaptive)
grid representations and meshfree representations
\citep{Saha1997,Abdelsalam1998,Abdelsalam1998b,Diego2005,Birrer2015,Merten2016,Galan2022}. 
These methods strive to model the signal contained in lensed images in a data-agnostic way, in order to place better constraints on the morphology of the source brightness or the 
projected mass density of the lens. 
However, most free-form parametrization choices make the inference problem under-constrained, meaning that imposing a prior on the reconstructed parameters becomes essential to penalize unphysical solutions and avoid overfitting the data.

In the context of traditional likelihood-based modeling, there exists a number of commonly used priors for the inference of high dimensional representations of background sources. 
For example, the strategy to impose a quadratic-log prior for linear inversion of pixelated-source models was developed by \citet{Warren2003,Suyu2006}. Other methods include 
iteratively specified priors for shapelets \citep{Birrer2015,Birrer2018,Nightingale2018} and a sparsity prior for wavelets representations \citep{Galan2021}.

On the other hand, for lens mass reconstruction, the issue of specifying an appropriate prior is still unsolved. This subject has been studied extensively in the context of cluster-scale strong lensing \citep{Bartelmann1996,Seitz1998,Abdelsalam1998,Abdelsalam1998b,Bradac2005,Diego2005,Cacciato2006,Diego2007,Liesenborgs2006,Liesenborgs2007,Jee2007,Coe2008,Merten2009,Deb2012,Merten2016,Ghosh2020,Torres-Ballestros2022}. Free-form approaches in the context of strong galaxy-galaxy lenses have been comparatively less studied (see however \citet{Saha1997,Saha2004,Birrer2015,Coles2014,Galan2022}). 

Another major challenge for these models is the issue of optimizing or sampling these high dimensional posteriors. 
Given the non-linear nature of the model and the existence of multiple local optima, non-linear global optimizers and samplers are needed, which often results in extremely expensive computational procedures.
The high computational cost of these methods also limits the extent to which they can be thoroughly tested and validated to identify and characterize potential systematics.

Over the recent years, deep learning methods have proven extremely successful at accurate modeling of strong lensing systems \citep{Hezaveh2017,PerreaultLevasseur2017,Morningstar2018,Coogan2020,Park2021,Legin2021,Legin2022,Wagner-Carena2021,Schuldt2022,Wagner-Carena2022,Karchev2022,AnauMontel2022,Mishra-Sharma2022,Schuldt2022}.
More specifically, \citet{Morningstar2019} demonstrated that recurrent convolutional neural networks can implicitly learn complex prior distributions from their training data to successfully reconstruct pixelated undistorted images of strongly lensed sources, circumventing the need to explicitly specify a prior distribution over those parameters. Motivated by this, we propose a method that extends this framework to solve the full lensing problem and simultaneously reconstruct a pixelated mass map and a pixelated image of the undistorted background source.

The method we propose here is based on the Recurrent Inference Machine (RIM), originally developed by \citet{Putzky2017}. In its original version, this method proposed to solve inverse problems using a Recurrent Neural Network as a metalearner to learn the iterative process of the optimization of a likelihood. RIMs have been trained on a range of linear inverse problems both within and outside of astrophysics \citep{Lonning2019}. In \cite{Modi2021}, this method was generalized to non-linear inference problems while using a U-net architecture \citep{Ronneberger2015} to separate the dynamics of different scales. 

In the present paper, we leverage this framework to learn an optimization process over the highly non-convex strong lensing likelihood, and implicitly learn a data-driven prior, which allows for the reconstruction of complex mass distributions representative of realistic galaxies taken from the IllustrisTNG \citep{Nelson2018} hydrodynamical simulations.
We also introduce a fine-tuning procedure, which allows us to directly exploit the prior encoded in the neural network parameters in order to further optimize the posterior down to noise levels. We apply this to the reconstruction of high signal-to-noise galaxy-galaxy lensing systems simulated using IllustrisTNG \citep{Nelson2018} projected density maps and background galaxy images collected from the COSMOS survey \citep{Koekemoer2007,Scoville2007}.

The paper is organised as follows. Section \ref{sec:methods} details 
the inference pipeline. In Section \ref{sec:data}, we present the data production and preprocessing for the training of the RIM and the generative models used in this paper. In Section \ref{sec:training}, 
we report on the training strategies used. 
In Section \ref{sec:results},
we discuss our results on a test set of gravitational lenses. We conclude in Section 
\ref{sec:conclusion}.

\section{Methods}\label{sec:methods}
Our goal is to predict pixelated maps of both the undistorted image of the background source and the projected density in the foreground lens from noisy lensed images. Our model consists of a Recurrent Inference Machine that predicts these variables of interest. Training this model requires a large number of training data, which we produce using a 
Variational Autoencoder (VAE) trained on density maps from the IllustrisTNG simulation and background sources from the Cosmos dataset (Section \ref{sec:vae training}).  

In this section, we present the structure of the lensing inference problem and provide information about our analysis method.
We begin with a general introduction to maximum a posteriori (MAP) inference in Section \ref{sec:maximum a posteriori}.
We describe the lensing simulation pipeline in Section \ref{sec:forward model}.
In Section \ref{sec:rim}, we motivate the use of the Recurrent Inference Machine and describe its  computational graph.
The architecture of the neural network is described in Section 
\ref{sec:gradient model}.  Finally, we describe the fine-tuning procedure and the transfer learning technique applied to achieve noise-level reconstructions in Section \ref{sec:fine-tuning}.

\subsection{Maximum a posteriori inference}\label{sec:maximum a posteriori}

We consider the task of reconstructing a vector of parameters of interest $\mathbf{x} \in \mathcal{X}$ given a vector of noisy
observed data $\mathbf{y} \in \mathcal{Y}$, a known forward (or physical) model $F$, and an additive noise vector $\boldsymbol{\eta}$. 
In what follows, we assume this vector to be sampled from a Gaussian distribution with known covariance matrix $C$, 
such that we can write
\begin{equation}\label{eq:MainEquation} 
\begin{aligned}
        \mathbf{y} &= F(\mathbf{x}) + \boldsymbol{\eta}\, ;\\[2pt]
        \boldsymbol{\eta} &\sim \mathcal{N}(0, C)\, .
\end{aligned}
\end{equation} 
In our case study, $F$ is a many-to-one non-linear 
mapping between the parameter space $\mathcal{X}$ 
and the data space $\mathcal{Y}$. 
Finding physically allowed solutions for this ill-posed inverse problem requires strong priors. The maximum a posteriori (MAP) solution maximizes the product of the likelihood $p(\mathbf{y} \mid \mathbf{x})$ and the prior $p(\mathbf{x})$:
\begin{equation}\label{eq:Posterior} 
        \hat{\mathbf{x}}_{\mathrm{MAP}} = \underset{\mathbf{x} \in \mathcal{X}}{ \mathrm{argmax}}\,\,
        \log p(\mathbf{y} \mid \mathbf{x}) + \log p(\mathbf{x})\, .
\end{equation} 
Assuming a Gaussian noise model for $\boldsymbol{\eta}$, the log-likelihood can be written analytically as
\begin{equation}\label{eq:Likelihood} 
        \log p(\mathbf{y} \mid \mathbf{x}) \propto -
        \big(\mathbf{y} - F(\mathbf{x})\big)^{T} C^{-1} \big(\mathbf{y} - F(\mathbf{x})\big)\, .
\end{equation} 
The prior distribution, however, is problem-dependent and encodes
expert knowledge of the model domain. As such, it is typically harder to write explicitly. 

\subsection{The Forward Model}\label{sec:forward model}

The forward model, $F$, is a simulation pipeline that receives a map of the surface brightness in the background source and a map of the projected density in the foreground lens to produce distorted images of background galaxies.
This pipeline uses ray tracing to calculate the deflection angles, $\bm{\alpha}$, and maps the observed coordinates, $\bm{\theta}$, into the coordinates of the background plane, $\bm{\beta}$, through the lens equation
\begin{equation}\label{eq:LensEquation}
        \bm{\beta} = \bm{\theta} - \bm{\alpha}(\bm{\theta})\, .
\end{equation}
The deflection angles are obtained using the projected surface 
density field $\kappa$ --- also referred to as convergence --- through the integral
\begin{equation}\label{eq:alpha}
        \bm{\alpha}(\boldsymbol{\theta}) = 
        \frac{1}{\pi} \int_{\mathbb{R}^2}
        \kappa(\boldsymbol{\theta}') 
        \frac{\boldsymbol{\theta}
        - \boldsymbol{\theta}'}{\lVert \boldsymbol{\theta} - 
        \boldsymbol{\theta}' \rVert^2}
        d^2\boldsymbol{\theta}'\, .
\end{equation}
Since we also use a discrete representation for the convergence, 
we approximate this integral by a discrete global convolution. Taking 
advantage of the convolution theorem, this operation 
can be computed in near-linear time using 
the Fast Fourier Transform algorithm (FFT). 

Assuming the observation 
has $M^2$ pixels, the convolution kernel 
would have $(2M + 1)^{2}$ pixels. 
Both the convergence map and the kernel are zero-padded 
to a size of $(4M+1)^{2}$ pixels in order to approximate a linear 
convolution and significantly reduce aliasing.

To produce simulated images, the intensity of an image pixel
is obtained through bilinear interpolation of the 
source brightness distribution at the coordinate $\boldsymbol{\beta}$.
A blurring operator --- convolution by a point spread function (PSF) --- is 
then applied to the lensed image to replicate the response of an optical system.
This operator is implemented as a GPU-accelerated matrix operation 
since the blurring kernels used in this paper have a significant proportion
of their energy distribution encircled inside a small pixel radius. Gaussian noise is then applied to the images, as described in more details in section \ref{sec:simulated observation}.

\subsection{Recurrent Inference Machine}\label{sec:rim}

Instead of handcrafting a prior distribution to solve the inverse problem (\ref{eq:MainEquation}), we build an inference pipeline with
a data-driven implicit prior encoded in a deep neural network architecture \citep{Bengio2009}.
The RIM \citep{Putzky2017} is a form of learnt
gradient-based inference algorithm, intended to solve inverse problems of the form  \eqref{eq:MainEquation}. This framework has mainly been applied in the context of linear under-constrained inverse problems --- i.e.\ where $F(\mathbf{x})$ can be represented as a matrix product $A\mathbf{x}$ --- for which the prior on the parameters $\mathbf{x}$, $p(\mathbf{x})$, is 
either intractable or hard to compute \citep{Morningstar2018,Morningstar2019,Lonning2019}. 
The use of the RIM to solve non-linear inverse problems was first investigated in \citep{Modi2021}.
In our case, the function representing the physical model $F$ encodes the lens equation (\ref{eq:LensEquation}), which is highly non-linear.

 The RIM is made up of a recurrent unit, which, given an observation $\mathbf{y}$, solves (\ref{eq:MainEquation}) for $\mathbf{x}$ through the governing equation
\begin{equation}\label{eq:RIM}
        \mathbf{\hat{x}}^{(t+1)} = \mathbf{\hat{x}}^{(t)} 
        + g_\varphi \big(\mathbf{\hat{x}}^{(t)},\, \mathbf{y},\, 
\grad_{\mathbf{\hat{x}^{(t)}}} \log p(\mathbf{y} \mid \mathbf{\hat{x}}^{(t)})\big)\, ,
\end{equation}
where $\mathbf{\hat{x}}^{(t)}$ is the estimate of the parameters of interest at time $t$ of the recursion (here, the pixel values of the image of the undistorted background source and of the density field $\kappa$) and $g_\varphi$ is a neural network. In the text, we will often use the shorthand notation $\grad_{\mathbf{y} \mid \mathbf{x}}$ to refer to $\grad_{\mathbf{x}} \log p(\mathbf{y} \mid \mathbf{x})$, the gradient of the likelihood evaluated at $\mathbf{x}$.
By minimizing a weighted mean squared loss backpropagated 
through time, 
\begin{equation}\label{eq:Loss}
		\mathcal{L}_\varphi(\mathbf{x}, \mathbf{y}) = 
		\frac{1}{T}\sum_{t=1}^{T}\sum_{i=1}^{M} \mathbf{w}_i (\mathbf{\hat{x}}^{(t)}_i - \mathbf{x}_i)^2\, ,
\end{equation} 
where $T$ is the total number of time steps in the recursion, the index $i$ labels the pixels of the reconstructions, $\mathbf{w}_i$ is the per-pixel weight, and $M$ is the total number of pixels in the reconstructions, the neural network $g_\varphi$ learns to optimize the parameters $\mathbf{x}$ given a likelihood function. 
The converged parameters of the neural network given the training set $\mathcal{D}$, 
$\varphi^{\star}_{\mathcal{D}}$, are those that minimize the cost --- or empirical risk ---
which is defined as the 
expectation of the loss over $\mathcal{D}$
\begin{equation}\label{eq:Cost} 
		\varphi^{\star}_{\mathcal{D}} = \underset{\varphi}{\text{argmin}}\,\,
		\mathbb{E}_{(\mathbf{x},\mathbf{y}) \sim \mathcal{D}}\big[
		\mathcal{L}_\varphi(\mathbf{x}, \mathbf{y}) \big].
\end{equation} 
Unlike previous works 
\citep{Andrychowicz2016,Putzky2017,Morningstar2018,Morningstar2019,Lonning2019}, 
the data vector $\mathbf{y}$ containing the observations 
is fed to the neural network in order to learn 
a better initialization of the parameters, $\mathbf{x}^{(0)} = g_\varphi(0, \mathbf{y}, 0)$, 
in addition to their optimization process. 
 Empirically, we found that this significantly improves the performance 
of the model for our problem and avoids situations where 
the model would get stuck in local minima at 
test time due to poor initialization. 
\begin{figure}[t!]
        \centering
		\includegraphics[width=\linewidth]{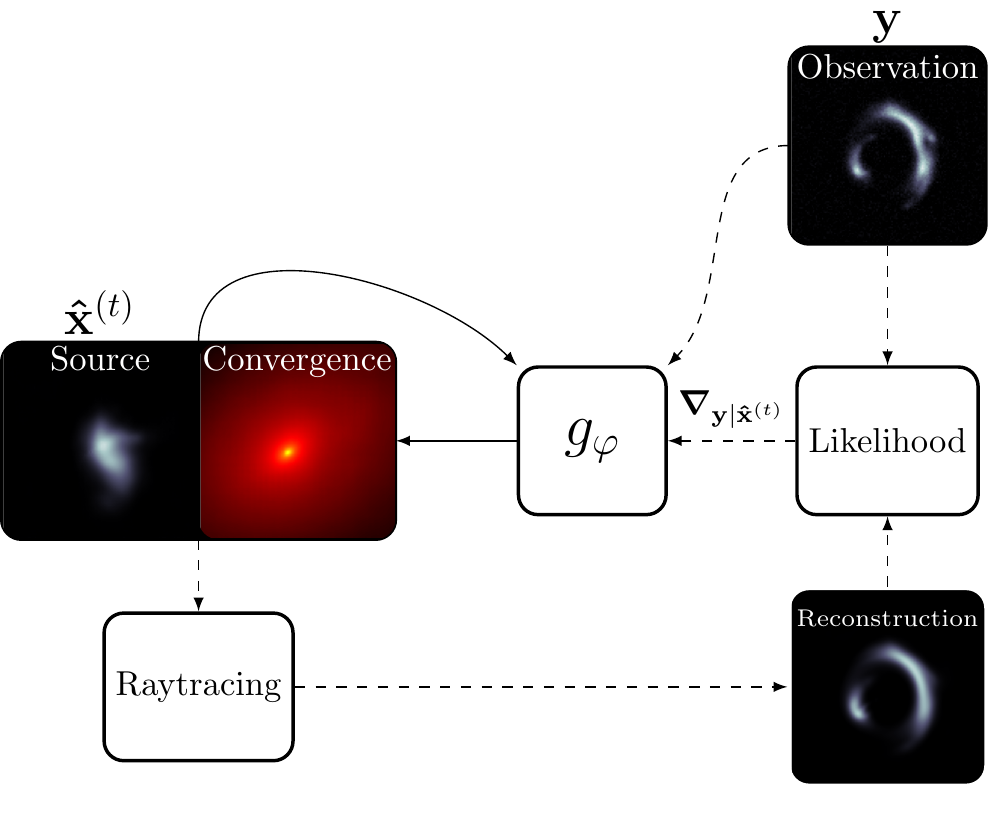}
        \caption{Rolled computational graph of the RIM. Dashed arrows represent operations not recorded for BPTT.}
        \label{fig:rolled graph}
\end{figure}

We follow previous works in setting a uniform weight over the time 
steps ($\mathbf{w}^{(t)} = \frac{\mathbf{w}}{T}$). 
The choice of the pixel weights $\mathbf{w}_i$ is informed
by our empirical observations when training the network. Details are reported in appendix \ref{ap:rim training and opt}.

In Figure \ref{fig:rolled graph}, we show the rolled computational graph of the 
RIM. During training of the neural network $g_\varphi$, operations along the solid arrows are being 
recorded for backpropagation through time. 
The recording is stopped along the dashed arrow since these operations 
are part of the forward modelling process and contain no trainable parameters.

\begin{figure*}[ht!]
        \centering
        \includegraphics[width=\textwidth]{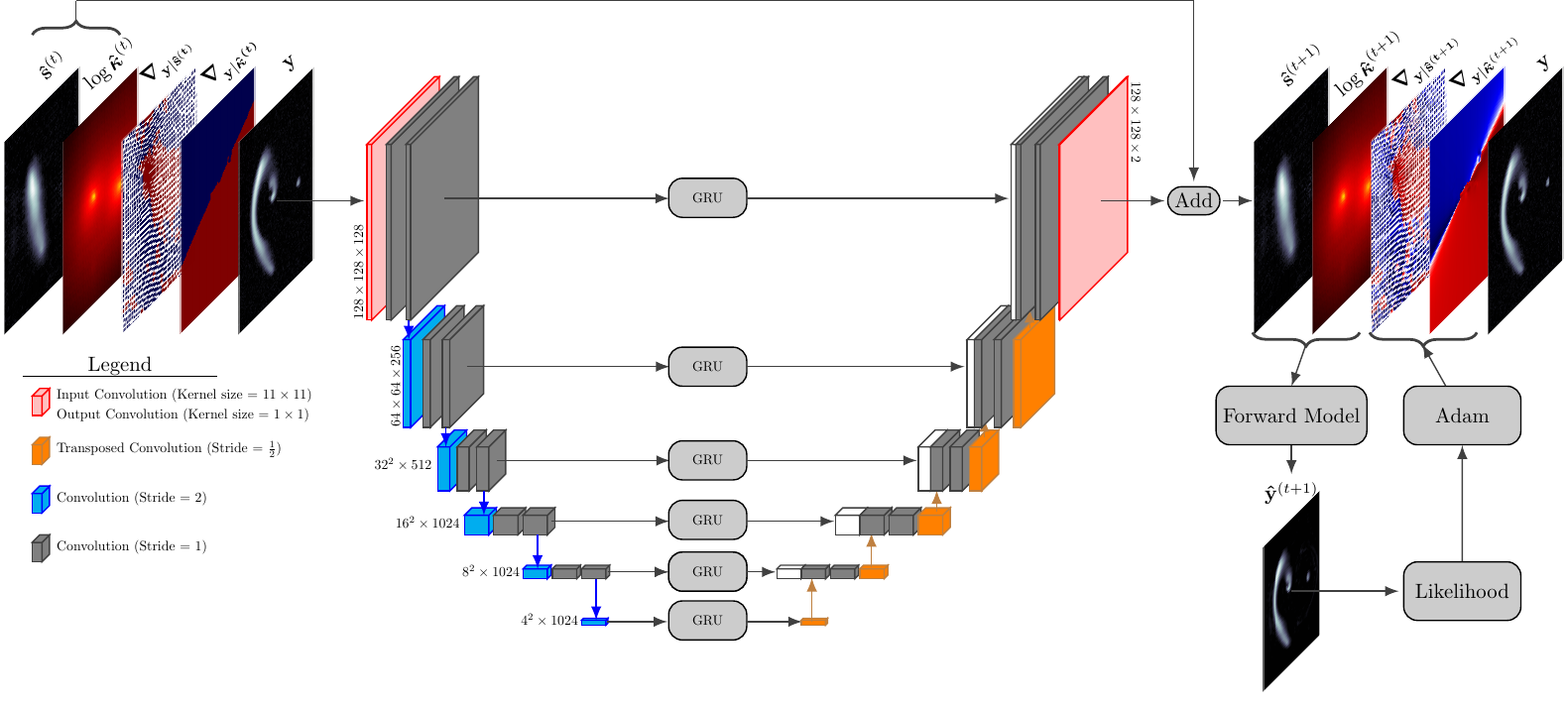}
        \caption{
A single time step of the unrolled computation graph of the RIM.
GRU units are placed in the skip connections to guide the 
reconstruction of the source and convergence. A schematic of the steps to compute 
the likelihood gradients is shown in the bottom right of the figure, including the 
Adam processing step of the likelihood gradient. See appendix \ref{ap:rim training and opt} for 
more details.}
        \label{fig:unet}
\end{figure*}

The gradient of the likelihood is computed using automatic differentiation. Following 
\citep{Modi2021}, we preprocess the gradients using the Adam algorithm \citep{Kingma2014}. 
For clarity, we only illustrate this step in Figure \ref{fig:unet}. 

\subsection{The Neural Network}\label{sec:gradient model}

The neural network architecture is illustrated in Figure \ref{fig:unet}, which shows 
a single time step of the unrolled computation graph of the RIM.
We use a U-net \citep{Ronneberger2015} architecture 
with Gated Recurrent Units \citep[GRU:][]{Cho2014} placed in each skip connection. 

Each GRU cell has its own memory tensor that is updated through time at each iteration of 
equation \ref{eq:RIM}. The shape of a memory tensor is set to match the
feature tensor fed into it from the parent layer in the network graph. 
Instead of learning a compressed representation like in the hourglass
architecture (or autoencoder), the U-net architecture naturally separates the spatial 
frequency components of the signal into its levels (neural network layers at the same height in 
Figure \ref{fig:unet}). The first level generally encodes 
high frequency features while the lower (or deeper) levels encode low frequency features (due to downsampling of the feature maps). 
Adding an independent memory unit 
at each level preserves this property.

Convolutional layers with a stride of 2 are used for downsampling and 
stride of $\frac{1}{2}$ for upsampling of the feature maps
(identified in blue and orange respectively in figure \ref{fig:unet}). Half-stride convolutions are implemented in practice with the transposed convolution layers from \texttt{Tensorflow} \citep{tensorflow}.
Most layers use a kernel size of $3\times3$, except the first and last layer. 
The first layer has 
larger receptive field ($11\times11$) in order to capture more details in the input tensor. 
The last layer has kernels of size $1\times 1$. 
A $\tanh$ 
activation function is used 
for each convolutional layer, including strided convolutions, except for the output 
layer. The U-net outputs an image tensor with two channels, one dedicated for the update of the source 
and the other for the update of the convergence (see figure \ref{fig:unet}). 



\subsection{Fine-Tuning}\label{sec:fine-tuning}

\subsubsection{Objective function}
Once trained, the RIM produces a baseline (point)
estimate of the parameters $\mathbf{x}$ given a noisy observation $\mathbf{y}$, a PSF and a noise covariance matrix. 
We now concern ourselves with a strategy to improve 
this estimate. 
This is important for observations with high SNR, for which the estimate
must be very accurate to model all the fine features present in the arcs.
The metric for the goodness of fit 
is the reduced chi squared $\chi^2_\nu = \frac{\chi^2}{\nu}$, 
where $\nu$ is the total number of degrees of freedom which here corresponds to 
the total number of pixels in $\mathbf{y}$.
Generally, our goal will be to reach $\chi^2_\nu = 1$, or equivalently $|\chi^2 - \nu| = 0$, 
which suggests that the RIM's estimate has modeled all the signal 
to be recovered from the observations. 
We note that this metric overestimates the 
number of degrees of freedom, which cannot be computed reliably for our particular model choice 
(pixelated source and convergence maps). In practice, the $\chi^2_{\nu}$ might be biased low.

To improve this figue of merit, we can optimize the log-likelihood directly w.r.t.\ the network weights given an appropriate prior on those weights (to avoid forgetting the implicit priors that have been learned during training, see section \ref{sec:transferlearning}). The new objective function is given by
\begin{equation}\label{eq:MAP} 
        \hat{\varphi}_{\mathrm{MAP}} = \underset{\varphi}{\mathrm{argmax}}\,\, 
        \frac{1}{T}\sum_{t=1}^{T} \log p(\mathbf{y} \mid \mathbf{\hat{x}}^{(t)}) + \log p(\varphi) \, ,
\end{equation} 
where $\varphi$ are the network weights, $\log p(\mathbf{y} \mid \mathbf{\hat{x}}^{(t)})$ is the log-likelihood, and $\log p(\varphi) $ is the log prior over the network weights.
Unlike the loss in equation \eqref{eq:Loss}, this objective function makes no use of labels 
($\mathbf{x}$). 
This allows us to use equation \eqref{eq:MAP} at test time in order to fine-tune the RIM's weights to a specific test example. 

\subsubsection{Transfer Learning}
\label{sec:transferlearning}
We now address the issue of transferring knowledge from the training task defined by the loss function in equation \eqref{eq:Cost}, 
to a test task specific to an observation, as defined by the loss given in equation \eqref{eq:MAP}.
The reader might refer to reviews on transfer learning \citep{Pan2010,Zhuang2019} 
for a broad overview of the field. The strategy we outline falls within 
the category of inductive transfer learning.



\begin{figure}[tb!]
       \centering 
       \includegraphics[width=\linewidth]{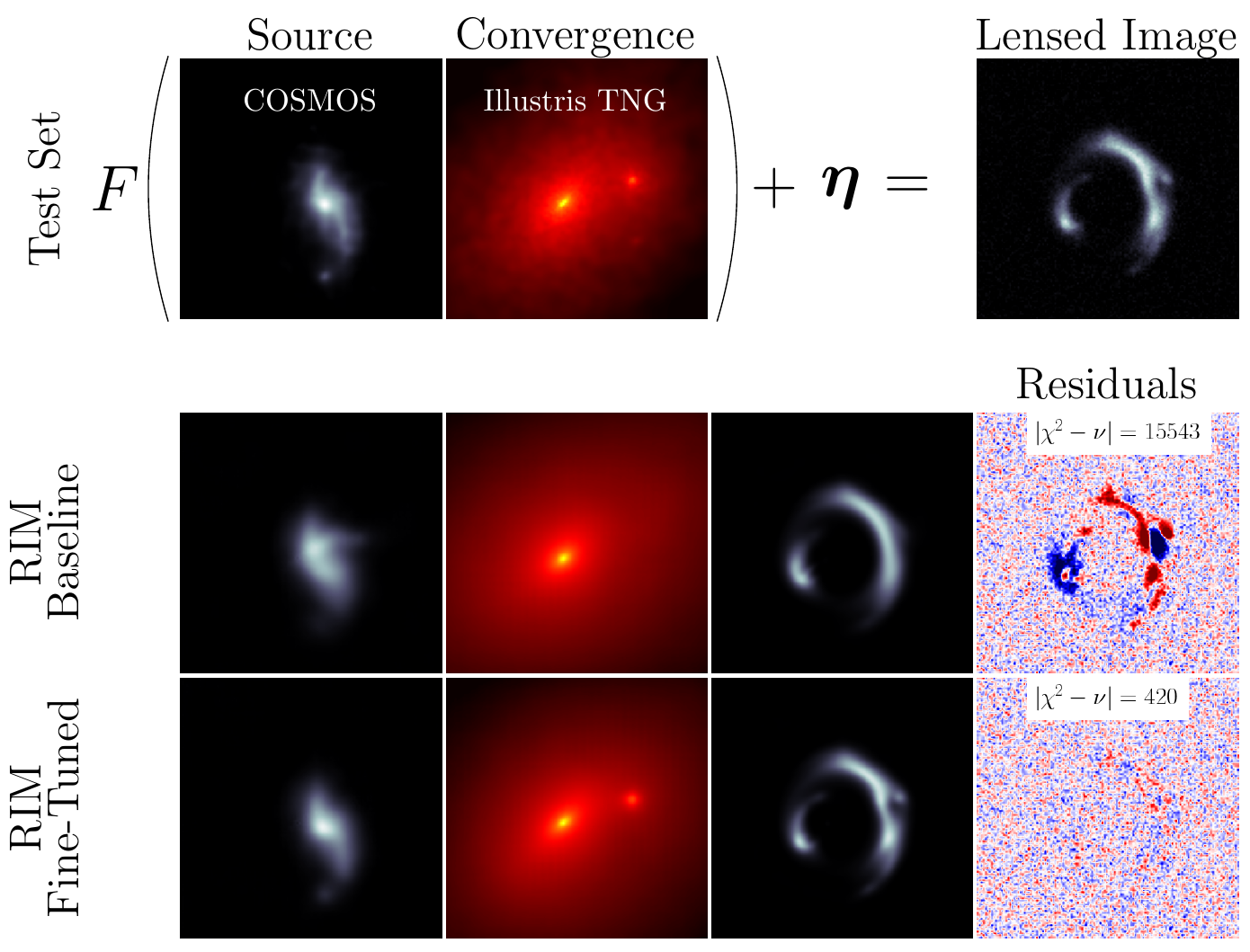}
       \caption{Example of a simulated lensed image in the test set that 
exhibits a large deflection in its eastern arc which indicates the presence of a massive object
 --- in this case a dark matter subhalo. The fine-tuning procedure is able to recover 
this subhalo because of its strong signal in the lensed image and reduces the residuals 
to noise level.}
       \label{fig:main figure}
\end{figure}

Optimizing the log-likelihood alone without a prior term over the weights (i.e.~just the first term from the r.h.s.~in \eqref{eq:MAP}) by initializing the weights at $\varphi^\star_\mathcal{D}$ is not strong 
enough to preserve the knowledge learned from the training task. 
This has long been observed in the literature and was coined as the 
catastrophic interference phenomenon in 
connectionist networks \citep{McCloskey1989,Ratcliff1990}.
In summary, a sequential learning problem exhibits catastrophic 
forgetting of old knowledge when confronted with new examples (possibly 
from a different distribution or process), in a manner 
\begin{enumerate}
        \item \label{cf:steps} proportional to the amount of learning;
        \item \label{cf:weights} strongly dependent on the disruption of the parameters
                involved in representing the old knowledge.
\end{enumerate}
While introducing an early stopping condition could 
potentially alleviate the former issue, the latter could still remain a problem.

We therefore follow the work of \citet{Kirkpatrick2016} to define a prior distribution
over $\varphi$ that address this issue
\begin{equation}\label{eq:Prior} 
        \log p(\varphi) \propto -\frac{\lambda}{2}\sum_{j} \mathrm{diag}(\mathcal{I}(\varphi_{\mathcal{D}}^{\star}))_{j} 
        (\varphi_j - [\varphi^{\star}_{\mathcal{D}}]_{j})^{2}\, .
\end{equation} 
where $\mathrm{diag}(\mathcal{I}(\varphi_{\mathcal{D}}^{\star}))$ is the diagonal of the 
Fisher information matrix 
encoding the amount of information that  
some set of gravitational lensing systems from 
the training set, and similar to the observed 
test task, carries about the baseline RIM weights $\varphi_{\mathcal{D}}^{\star}$ 
--- the parameters that minimize the empirical risk (equation \ref{eq:Cost}).
We can also understand this prior using the
Cramér-Rao lower bound 
\citep{Rao1945,Cramer1946}.
The prior can thus be framed as a multivariate 
Gaussian distribution characterised by a diagonal covariance matrix with $\mathrm{diag}(\mathcal{I})$ as its inverse 
and by $\varphi^{\star}_{\mathcal{D}}$ as its first moment. 
Within this view, the  
Lagrange multiplier is 
tuning our estimated uncertainty about the neural network weights 
for the particular task at hand.  
We have included a derivation 
of this term in the appendix \ref{ap:ewc}.

\begin{figure}[t!]
        \centering
        \includegraphics[width=\linewidth]{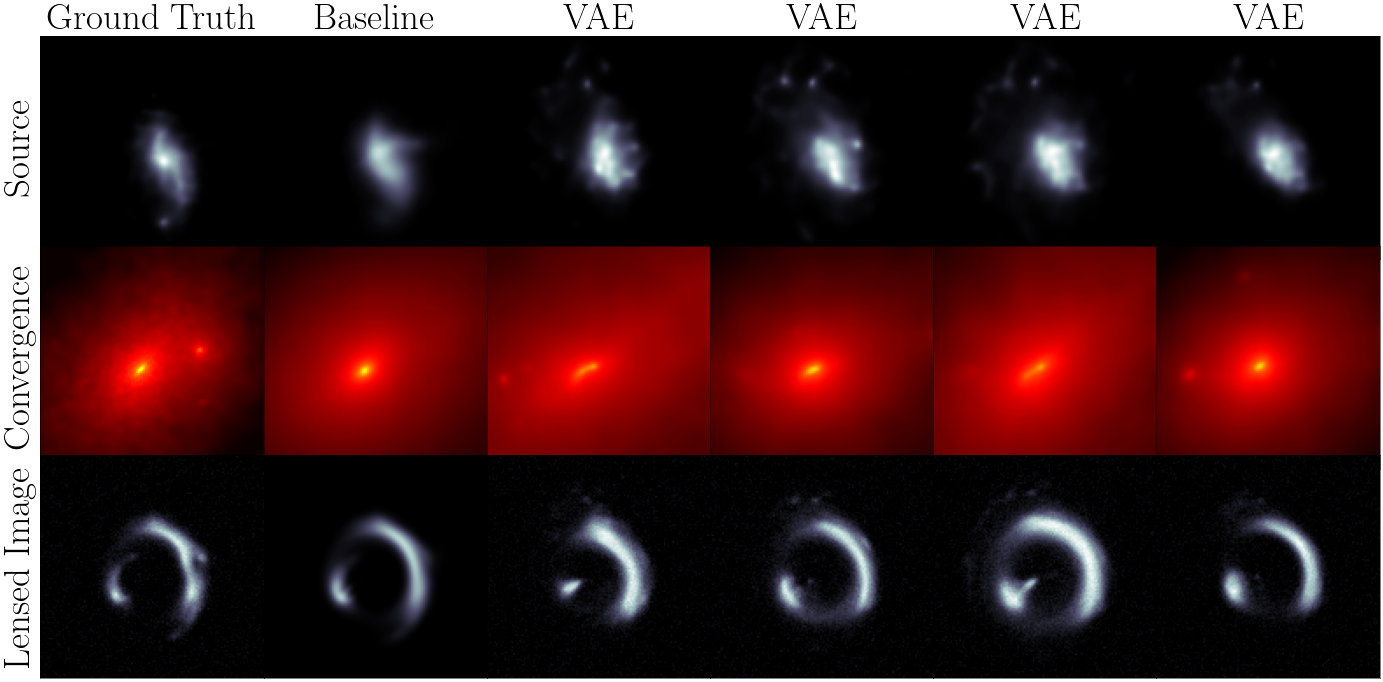}
        \caption{Examples similar to the test task, also shown in Figure \ref{fig:main figure}. The first column shows the ground truth used to simulate the lensed image. The second column shows the baseline prediction that is then encoded in the latent space of the VAE in order to sample the next 4 columns.
}
        \label{fig:vae fine-tuning}
\end{figure}

Examples are drawn from the set of training examples similar to the test task by sampling the latent space of two variational autoencoders (VAE) that model a distribution over the background sources and the convergence maps respectively (as described in Section \ref{sec:source} and \ref{sec:kappa}) near the baseline prediction of the RIM. In practice, we choose an isotropic Gaussian distribution centered around $\hat{\mathbf{z}}^{(T)}$ --- 
the latent code of the baseline prediction --- as a sampling distribution. While we leave the possibility of improving this choice to future work, it is sufficient for our goals. 
Figure \ref{fig:vae fine-tuning} illustrates examples of what is meant here by \textit{similar}. 

\section{Data}\label{sec:data}

\subsection{COSMOS}\label{sec:source}
The maps of surface brightness of background sources are taken from the \textit{Hubble Space Telescope} (\textit{HST}) 
Advanced Camera for Surveys Wide Field Channel COSMOS field \citep{Koekemoer2007,Scoville2007},
a $1.64\,\mathrm{deg}^{2}$ contiguous survey acquired in the F814W filter. 
A dataset of magnitude limited ($\mathrm{F814W} < 23.5$) deblended galaxy postage stamps \citep{Leauthaud2007} was compiled as 
part of the GREAT3 challenge \citep{Mandelbaum2014}. The data is 
publicly available \citep{Mandelbaum2012}, and the preprocessing is done through the open-source software 
\texttt{GALSIM} \citep{Rowe2015}. \par

We apply the 
\texttt{marginal} selection criteria (see the \texttt{COSMOSCatalog} class) and impose a flux per image
greater than $50\,\,\mathrm{photons}\,\,\mathrm{cm}^{-2}\,\mathrm{s}^{-1}$. 
This final set has a total of 13\,321 individual images.
Each image is saved as a postage stamp of $158^2$ pixels. 
We then subtract the background from each image, apply a random shift, rotate them by an angle multiple of $90^\circ$, crop them down to $128^{2}$ pixels, and finally normalize them to pixel intensities in the range $[0,1]$. We then train an autoencoder to denoise the galaxy images \citep{Vincent2008,Vincent2010}. More specifically, we use the informational bottleneck principle \citep{Tishby2000} to learn a lossy lower-dimensional representation of the data. For a generic CNN autoencoder, this amounts to learning a low-pass frequency filter on the COSMOS dataset. Indeed, CNNs are known to exhibit a spectral bias in their learning phase \citep{Rahaman2018}, which we exploit to our advantage in order to filter pixel noise from the galaxy surface brightness. Furthermore, using an expressive CNN autoencoder produces much fewer artifacts than a naive implementation of such a low-pass filter --- e.g.\ by masking Fourier modes. 

We split the galaxies into a training set (90\%) and a test set (10\%). 
The augmented training set (${\sim50\,000}$ images) is used to train a VAE, 
as described in Section \ref{sec:vae training}, and produce simulated observations to train the RIM.

\begin{figure}[t!]
        \centering
        \includegraphics[width=\linewidth]{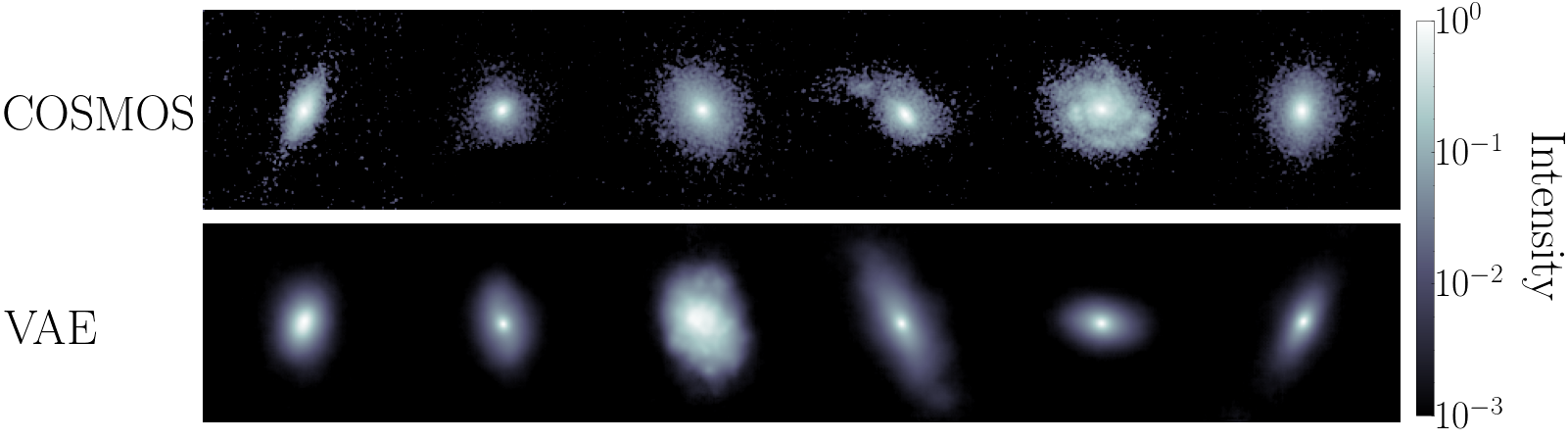}
        \caption{Examples of COSMOS galaxy images 
                (top row) and VAE generated samples (bottom row) used as labels in $\mathcal{D}$.}
        \label{fig:source}
\end{figure}

\subsection{IllustrisTNG}\label{sec:kappa}
\subsubsection{Smooth Particle Lensing}\label{sec:SPL}
To compute convergence maps from an N-body simulation, 
we use Kernel Density Estimation to produce smooth densities on a regular grid from discrete simulation particles.
This reduces the particle noise affecting all 
important lensing quantities. At the same time, the choice of the kernel size 
is important to preserve substructures in the 
lens that we might potentially be interested in. Following \citet{Aubert2007,Rau2013}, we use Gaussian smoothing with an adaptive 
kernel size determined by the distance of the 64\textsuperscript{th} nearest neighbours, $D_{64,i}$, of a given particle of 
mass $m_i$ and projected position $\mathbf{x}_i$
\begin{equation}\label{eq:Ksmooth}
\begin{aligned}
    \kappa(\mathbf{x}) &= \frac{1}{\Sigma_{\mathrm{crit}}} \sum_{i=1}^{N_{\mathrm{part}}}
        \frac{m_i}{2 \pi \hat{\ell}^2_i} 
        \exp \left(-\frac{1}{2} \frac{\lVert \mathbf{x} - \mathbf{x}_i\rVert^2}{\hat{\ell}_i^2}  \right) \\
    \hat{\ell}_i &= \sqrt{\frac{103}{1024}}D_{64,i}.
\end{aligned}
\end{equation}
The nearest neighbours are found by fitting a k-d tree ---  implemented in 
\texttt{scikit-learn} \citep{scikit-learn} --- 
to the $N_{\mathrm{part}}$  particles 
in a cylinder centered on the centre of mass of the halo of interest.
The critical surface density is defined as
\begin{equation}\label{eq:Scrit}
\Sigma_{\mathrm{crit}} = \frac{c^{2} D_s}{4 \pi G D_{\ell s} D_{\ell}}
\end{equation}
where $D_\ell$, $D_s$ and $D_{\ell s}$ are angular diameter distances to the lens, source and between the lens and the source respectively, 
$G$ is the gravitational constant, and $c$ the speed of light.

\subsubsection{Preprocessing}
The projected surface density maps (convergence) of lensing galaxies are made using the redshift $z=0$ snapshot  
of the IllustrisTNG-100 simulation \citep{Nelson2018} 
in order to produce physically realistic realizations of density maps containing dark and baryonic matter.
We select 1604 halos with the criteria that they have a total
dark matter mass of at least $9\times10^{11} M_{\odot}$. We then collect all 
dark matter, gas, stars and black holes particles from the data in the vicinity of the halo to 
create a smooth projected surface density map as prescribed in section \ref{sec:SPL}.

\begin{figure}[t!]
        \centering
        \includegraphics[width=\linewidth]{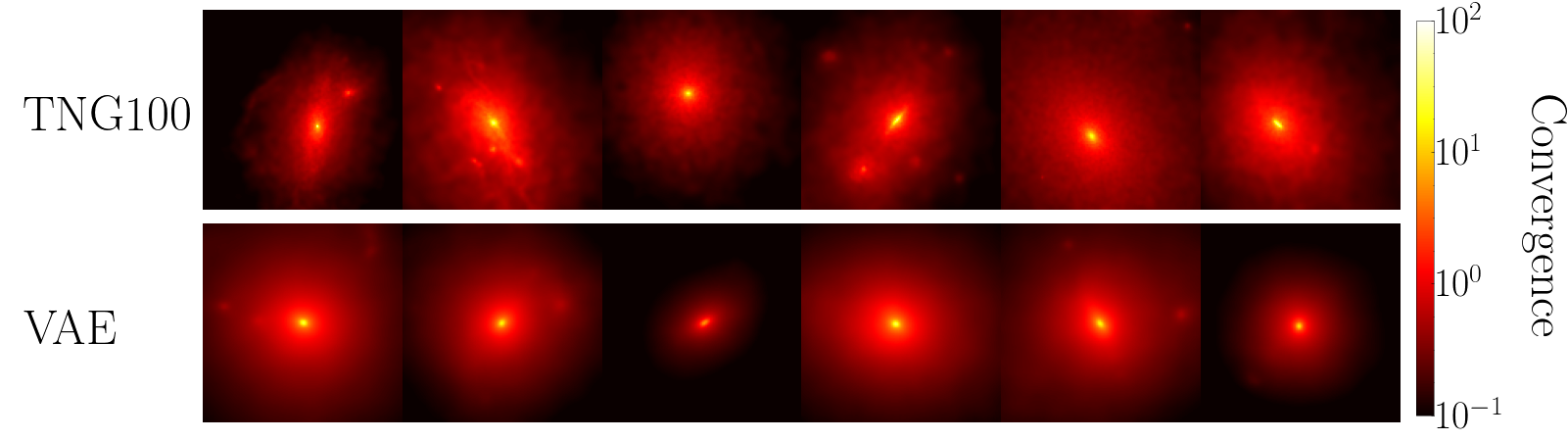}
        \caption{Examples of smoothed Illustris TNG100 convergence map (top row) 
        and VAE generated samples (bottom row) used as labels in $\mathcal{D}$.}
        \label{fig:kappa}
\end{figure}

We adopt the $\Lambda$CDM cosmology from 
\citet{PlanckCollaboration2018} with $h=0.68$ to compute 
angular diameter distances. We also fix the 
source redshift to $z_s=1.5$ and the deflector redshift to $z_\ell=0.5$. 
We note that changing the redshifts or the cosmology 
only amount in a rescaling of the $\kappa$ map by a global scalar. Thus, this choice does not change the generality of our method.
The smoothed density maps from equation \eqref{eq:Ksmooth} are 
rendered into a regular grid of $188^2$ pixels with a comoving field of view of $105\,\,\mathrm{kpc}/h$. 
To avoid edge effects in the pixelated maps, 
we include particles outside of the field of view in the sum of equation \eqref{eq:Ksmooth}.
\par
Before applying augmentations or considering different projections, our dataset of halos is split into a 
training set (90\%) and a test set (10\%), in order to make sure that the test set consists only 
of convergence maps unseen by the RIM during training.
We take 3 different projections ($xy$, $xz$ and $yz$) of each 3D particle 
distribution, which amounts to a dataset with a total of $4\,812$ individual convergence maps. 
Random rotations by an angle multiple of $90^{\circ}$ and random shifts to the pixel coordinates 
are applied to each image. The $\kappa$ maps are then rescaled by a random factor to change their 
estimated Einstein radius to the range 
$[0.5,\,2.5]$ arcseconds.
The Einstein radius is defined as
\begin{equation}\label{eq:ThetaE}
        \theta_E = \sqrt{\frac{4GM(\theta_E)}{c^ 2} \frac{D_{\ell s}}{D_\ell D_s}}
\end{equation} 
where $M(\theta_E)$ is the mass enclosed inside the Einstein radius. In practice, we estimate this quantity 
by summing over the mass of pixels with a value greater than the critical density ($\kappa > 1$). 
For data augmentation purposes, this procedure gives a good enough estimate of the lensed image separation resulting from a given $\kappa$ map. 
We test multiple scaling factors for each $\kappa$ map, then uniformly sample between those that produce an estimated 
Einstein radius within the 
desired range. This step is used to remove any bias in the Einstein radius that might come from the mass function 
of the simulation.

The final maps are cropped down to $128^2$ pixels.
Placed at a redshift $z_\ell=0.5$, a $\kappa$ map will thus span an angular field of view of $7.69''$ with 
a resolution similar to \textit{HST}. 
With these augmentation procedures, we create a total of $50\,000$ maps from the training split
to train a VAE, as described in Section \ref{sec:vae training}, and produce simulated observations to train the RIM.

\subsection{Simulated Observations}\label{sec:simulated observation}
With a given source map and convergence map, we apply the ray tracing simulation 
described in section \ref{sec:forward model} to produce a lensed image.

For each lensed image, we create a Gaussian PSF  
with a full width at half maximum (FWHM) 
randomly generated from a truncated normal distribution.
The support of the distribution is truncated below by the 
angular size of a single pixel and above by the angular size of 4 pixels. 
White noise with a standard deviation randomly generated from a truncated normal distribution 
is then added to the convolved lensed image to simulate noisy observations. These noise realizations result in SNRs between 
$10$ and $1000$. For simplicity, we define $\mathrm{SNR} = \frac{1}{\sigma}$. 
This definition is equivalent to the peak signal-to-noise ratio. 

To ensure that the images are representative of strongly lensed sources, we require a minimum flux magnification of 3. We
also make sure that most pixel coordinates in the image plane are mapped inside the 
source coordinate system through the lens equation \eqref{eq:LensEquation}. 

\begin{table}[htb!]
\centering
\caption{Physical model parameters.}
\label{tab:phys}
\begin{tabular}{ccc}
        Parameter &  Distribution/Value \\
        \hline \hline
        Lens redshift $z_\ell$ & $0.5$ \\
        Source redshift $z_s$ & $1.5$ \\
        Field of view ('') & $7.69$ \\
        Source field of view ('') & $3$ \\
        PSF FWHM ('') & $\mathcal{TN}(0.06,\, 0.3;\, 0.08,\, 0.05)$
        \footnote{We defined the parameters of the truncated normal in the order $\mathcal{TN}(a,\, b;\, \mu,\, \sigma)$, where $[a,\, b]$ defines the support of the distribution.} \\
        Noise amplitude $\sigma$ & $\mathcal{TN}(0.001,\, 0.1;\, 0.01,\,0.03)$\\
        \hline
\end{tabular}
\end{table}

In total, $400\,000$ training observations are simulated from random pairs of COSMOS sources 
and IllustrisTNG convergence maps in order to train the RIM. 
We create an additional $200\,000$ observations from pairs 
of COSMOS sources and pixelated SIE convergence maps. 
The parameters for these $\kappa$ maps are listed in table \ref{tab:sie}. 

\begin{table}[htb!]
\centering
\caption{SIE parameters.}
\label{tab:sie}
\begin{tabular}{ccc}
        Parameter &  Distribution \\
        \hline \hline
         Radial shift ('') & $\mathcal{U}(0, 0.1)$ \\
        Azimutal shift & $\mathcal{U}(0, 2\pi)$ \\
        Orientation & $\mathcal{U}(0, \pi)$ \\
        $\theta_E$ ('') & $\mathcal{U}(0.5, 2.5)$ \\
        Ellipticity & $\mathcal{U}(0, 0.6)$ \\
        \hline
\end{tabular}
\end{table}

We generate $1\,600\,000$ simulated observations from the VAE 
background sources and convergence maps as part of the training set. We apply some validation checks to each example in order to avoid configurations like a single image of the background source or an Einstein ring cropped by the field of view.

\section{Training}\label{sec:training}

\subsection{VAE}\label{sec:vae training}
Here, we describe the training of two VAEs that are used to produce density maps and images of unlensed background galaxies to train and test our inference model. For an introduction to VAEs we refer the reader to \citet{Kingma2019}.

As mentioned in \citet{Kingma2019}, direct optimisation 
of the ELBO loss can prove difficult because the reconstruction term $\log p_\theta (\mathbf{x} \mid \mathbf{z})$ 
is relatively weak compared to the Kullback Leibler (KL) divergence term. To alleviate this issue, 
we follow the work of \citet{Bowman2015} and \citet{Sonderby2016} in setting a warm-up 
schedule for the KL term, starting from $\beta=0.1$ up to $\beta_{\mathrm{max}}$. 

Usually, 
$\beta_{\mathrm{max}} = 1$ is considered optimal since it matches the original ELBO  
objective derived by \citet{Kingma2013}. 
However, we are more interested in the 
sharpness of our samples and accurate inference around small regions of the latent 
space for fine-tuning. Thus, setting $\beta_{\mathrm{max}} < 1$ allows us to increase 
the size of the information bottleneck (i.e.~latent space) of the VAE 
and improve the reconstruction cost of the model. 
This is a variant of the $\beta$-VAE \citep{Higgins2017}, where $\beta > 1$ was found 
to improve disentangling of the latent space \citep{Burgess2018}. 

The value for $\beta_\mathrm{max}$ and the steepness of the schedule 
are grid searched alongside the architecture for the VAE. 
These values are found in practice by 
manually looking at the quality of generated samples for different VAE 
hyperparameters. A similar method is explored and formalized in the 
InfoVAE framework \citep{Zhao2017}.

A notable element of the VAE architecture is the use of a fully connected
layer to reshape the features of the convolutional layer into the chosen 
latent space dimension. Following the work of \citet{Lanusse2021}, we introduce 
an $\ell_{2}$ penalty between the input and output of the bottleneck 
dense layers to encourage an identity mapping. This regularisation 
term is slowly removed during training.

\subsection{RIM}\label{sec:rimtraining}

The architecture of the neural network is grid searched on 
a smaller dataset ($\lesssim 10\,000$ examples) 
in order to quickly identify a small set
of valid hyperparameters. The best hyperparameters are 
then identified using a two-stage training process. 
In the first stage, we train 24 different architectures from this small hyperparameter set for approximately 4 days (wall time using a single Nvidia A100 GPU). 
Different architectures have a training time much longer than others, which 
is factored in the architecture selection process. For example, adding more time steps ($T$) to the recurrent relation \eqref{eq:RIM} 
yields better generalisation on the test set, but this 
comes at great costs to training time until convergence. Thus, $T < 10$ is preferred.

Following this first stage, 4 architectures are deemed efficient enough 
to be trained for an additional 6 days. 
We only report the results for the best architectures out of these 4.
\begin{figure*}[ht!]
        \centering
        \includegraphics[width=0.9\linewidth]{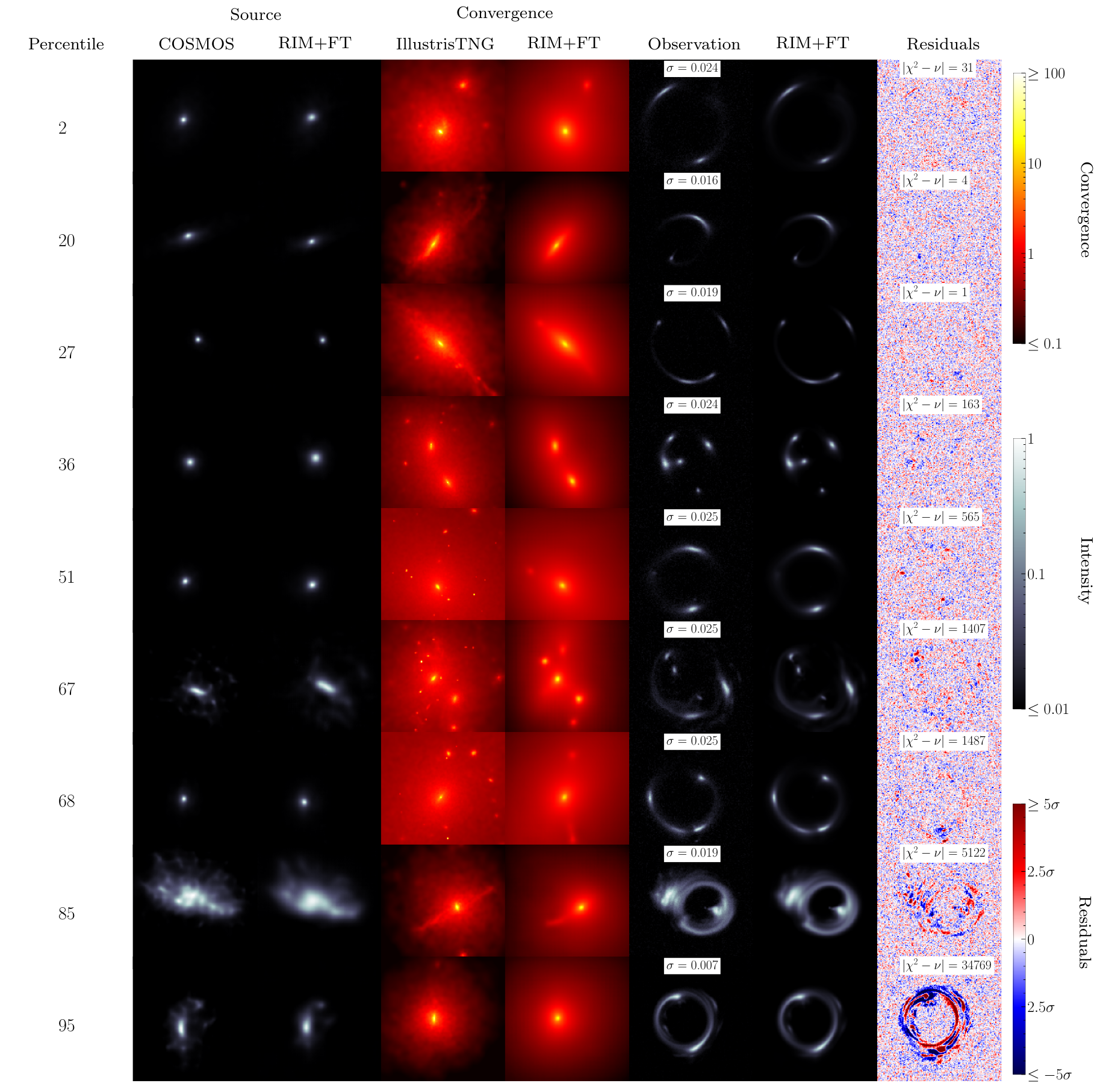}
        \caption{
                Sample of the fine-tuned RIM reconstructions 
                on a test set of 3000 examples. 
                Examples are ordered from the best $\chi^2$ (top) to the worst (bottom). 
                The percentile rank of each example is in the leftmost column. 
                The last example 
        shown has SNR above the threshold defined in Figure \ref{fig:chi squared vs noise}.}
        \label{fig:main result}
        \vspace{-1.5pt} 
\end{figure*}

\begin{figure*}[thb!]
        \centering
        \includegraphics[width=0.8\textwidth]{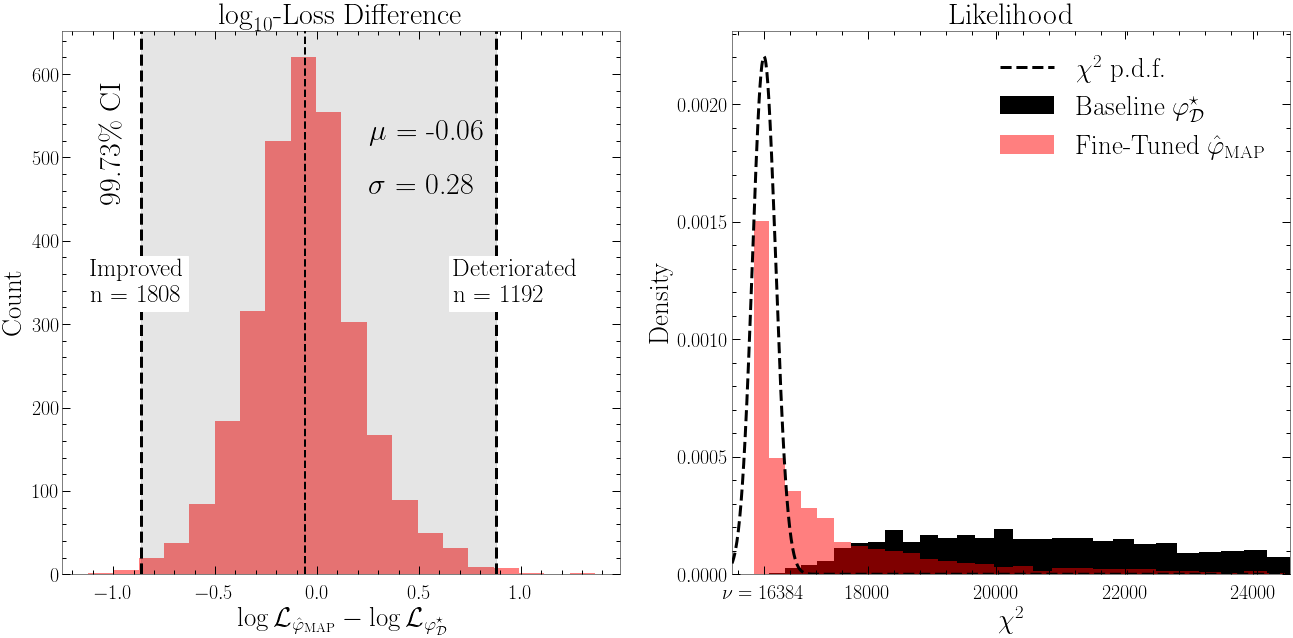}
        \caption{Distribution of the goodness of fit for the baseline and fine-tuned network (right panel), as well as log-loss difference between the two network for a given example in the test set (left panel).
}
        \label{fig:loss and chi squared}
\end{figure*}

Each reconstruction is performed by fine-tuning the baseline model 
on a test task composed of an observation, a PSF, and a noise covariance.
In practice, fine-tuning predictions on the test set of $3\,000$ examples can be accomplished in parallel so as to be done in at most a few days by spreading the computation on $\sim 10$ Nvidia A100 GPUs. Each reconstruction uses at most 2000 steps, correspondling to approximately $20$ minutes (wall-time) per reconstruction. Early stopping is applied when the $\chi^2$ reaches noise level. The hyperparameters for this procedure are reported in Table \ref{tab:fine-tuning hparams}.

\begin{table}[htb!]
        \centering
        \caption{Hyperparameters for fine-tuning the RIM.}
        \label{tab:fine-tuning hparams}
        \begin{tabular}{cc}
                Parameter & Value \\\hline\hline
                Optimizer & RMSProp \\
                Learning rate & $10^{-6}$\\
                Maximum number of steps & $2\,000$\\
                $\lambda$ & $2\times 10^{5}$\\
                $\ell_2$ & 0 \\
                Number of samples from VAE & 200 \\
                Latent space distribution & $\mathcal{N}(\mathbf{z}^{(T)}, \sigma=0.3)$
                \footnote{$\mathbf{z}^{(T)}$ is the latent code of the RIM baseline source or convergence.}\\
                \hline
        \end{tabular}
\end{table}

\section{Results}\label{sec:results}

In this section, we present the performance of our model 
on the held out test set. A sample of 3000 reconstruction 
problems is generated from the held-out \textit{HST} and IllustrisTNG data 
with noise levels and PSFs similar to the training set.

\subsection{Goodness of Fit}
Figure \ref{fig:main result} shows a sample of reconstructions for high SNR data with a wide range of lensing configurations from the test set.
We select examples representative of all levels of reconstruction performance (covering the entire range of goodness of fit) for data with  complex structures in their convergence map to showcase the expressivity of the approach. 
We also show a randomly selected sample from the test set in Figure \ref{fig:random sample}.

\begin{table}[]
    \centering
    \caption{$\log_{10}$-normal moments of the loss on the test set}
    \label{tab:loss}
    \begin{tabular}{ccc}
        \hline
          Model  & $\mu(\log \mathcal{L}_\varphi)$ & $\sigma(\log \mathcal{L}_\varphi)$ \\
        \hline \hline
        Baseline ($\varphi_{\mathcal{D}}^\star)$ &  -1.96 & 0.36 \\
        Fine-tuned ($\hat{\varphi}_{\mathrm{MAP}}$) & -2.02 & 0.37 \\\hline
    \end{tabular}
\end{table}

Figure \ref{fig:loss and chi squared} shows a comparison between 
the goodness of fit of the baseline model and the fine-tuned prediction. 
Since we empirically observe that the distribution of the loss on the test set (and the training set) follows a log-normal distribution, we find that it is more informative to look at the $\log$-loss 
distribution to extract information about the fine-tuning procedure. 
The left panel of Figure \ref{fig:loss and chi squared} 
shows the distribution of the log-loss difference between the fine-tuned prediction and the baseline model. This distribution shows that the fine-tuning procedure loss is constrained within $\sim 1$ order of magnitude of the original loss with a probability $>99.73\,\%$. We find that the log-loss difference has a scatter of $\sigma = 0.28$, which is smaller than the scatter of the baseline log-loss over the entire test set $\sigma(\log \mathcal{L}_{\varphi^\star_{\mathcal{D}}}) = 0.36$ reported in Table \ref{tab:loss}.
We note that the loss is not optimized during fine-tuning, still we notice that the fine-tuning procedure does not significantly deteriorate or improve the loss of the baseline prediction on average. We report the first 2 moments of the loss log-normal distribution for the baseline and the fine-tuned reconstructions in Table \ref{tab:loss} in order to explicitly compare them. As can be seen in this table, there is no significant difference between the two distributions. This statement can be proven for the measured mean values --- $\mu(\log \mathcal{L}_{\hat{\varphi}_{\mathrm{MAP}}}) = \mu(\log \mathcal{L}_{\varphi^{\star}_{\mathcal{D}}}) $ --- using the two-sided normal p-value test \citep{Casella2001}, which we find satisfy the null hypothesis with $p=0.87$ ($Z = -0.16$). All those observations support our claim that EWC regularisation preserves the prior learned during pretraining, or at least that it preserves the surrogate measures of the prior we reported. 

\begin{figure}[t!]
        \centering
        \includegraphics[width=\columnwidth]{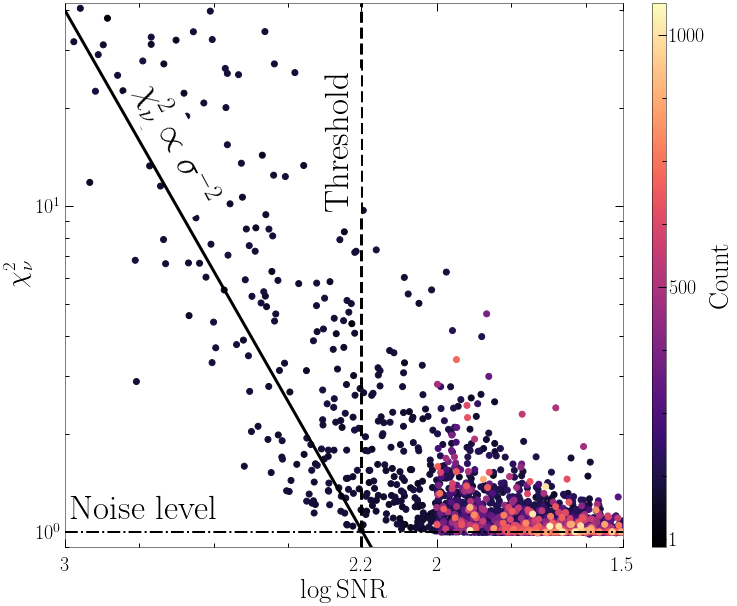}
        \caption{Goodness of fit as a function of SNR shows a threshold 
        behavior where our method reaches its limit.}
        \label{fig:chi squared vs noise}
\end{figure}

\begin{figure*}[t!]
        \centering
        \tikzset{font={\fontsize{8pt}{12}\selectfont}}
        \begin{tikzpicture}
                \node at (0, 0) {\includegraphics[width=0.9\linewidth]{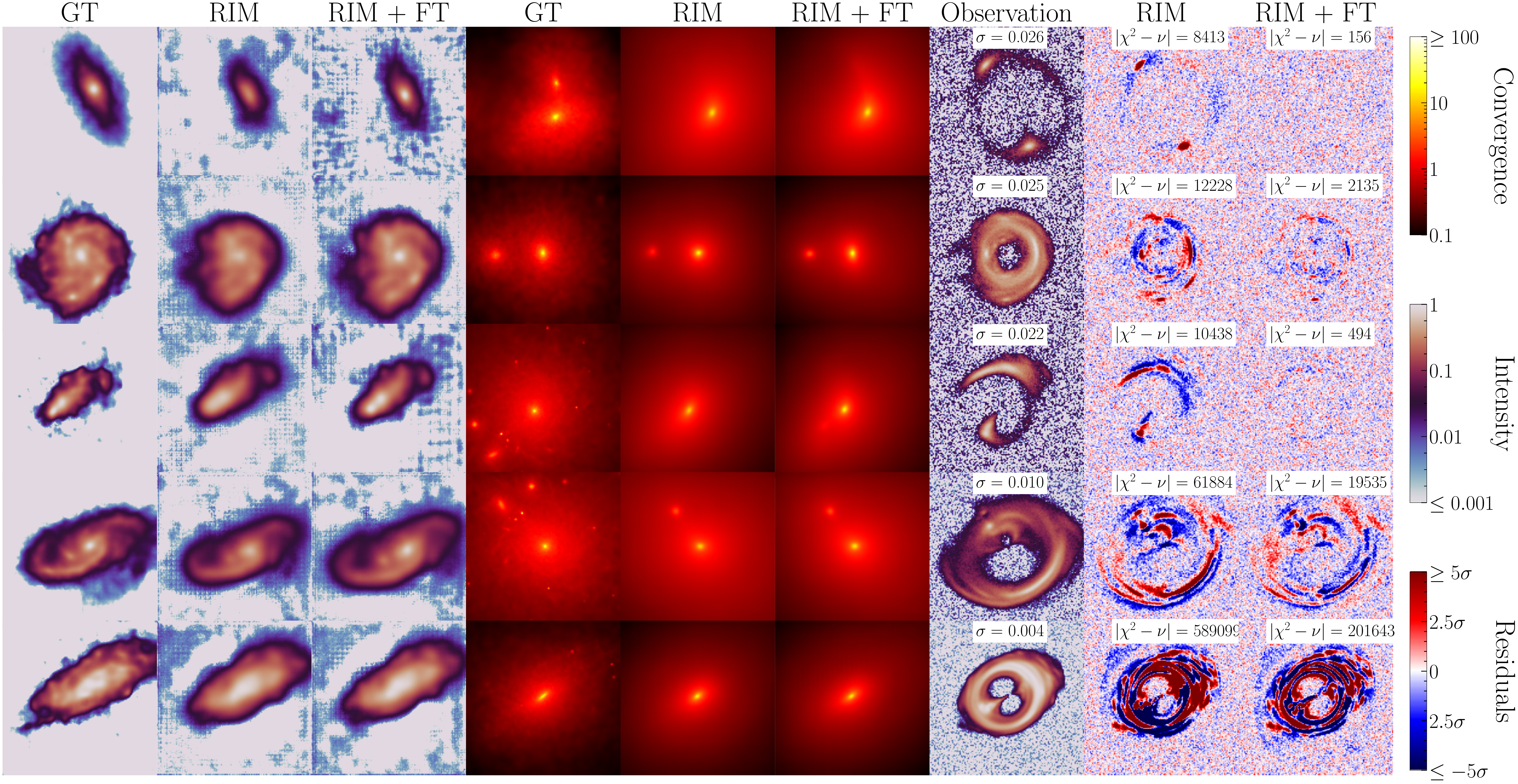}};
                \draw[-latex] (-8.5, 3.5) -- (-8.5, -3.5) node[midway, above, rotate=90] {Increasing SNR};
                \node at (-5.5, 4.5) {\strut Source};
                \node at (-0.75, 4.5) {\strut Convergence};
                \node at (5, 4.5) {\strut Residuals};
        \end{tikzpicture}
        \caption{
        Comparison between baseline (RIM) and fine-tuned (RIM+FT) reconstructions for gravitational lensing systems from the test set (GT).
        From top to bottom, we increase SNR. 
        }
        \label{fig:increasing SNR}
\end{figure*}
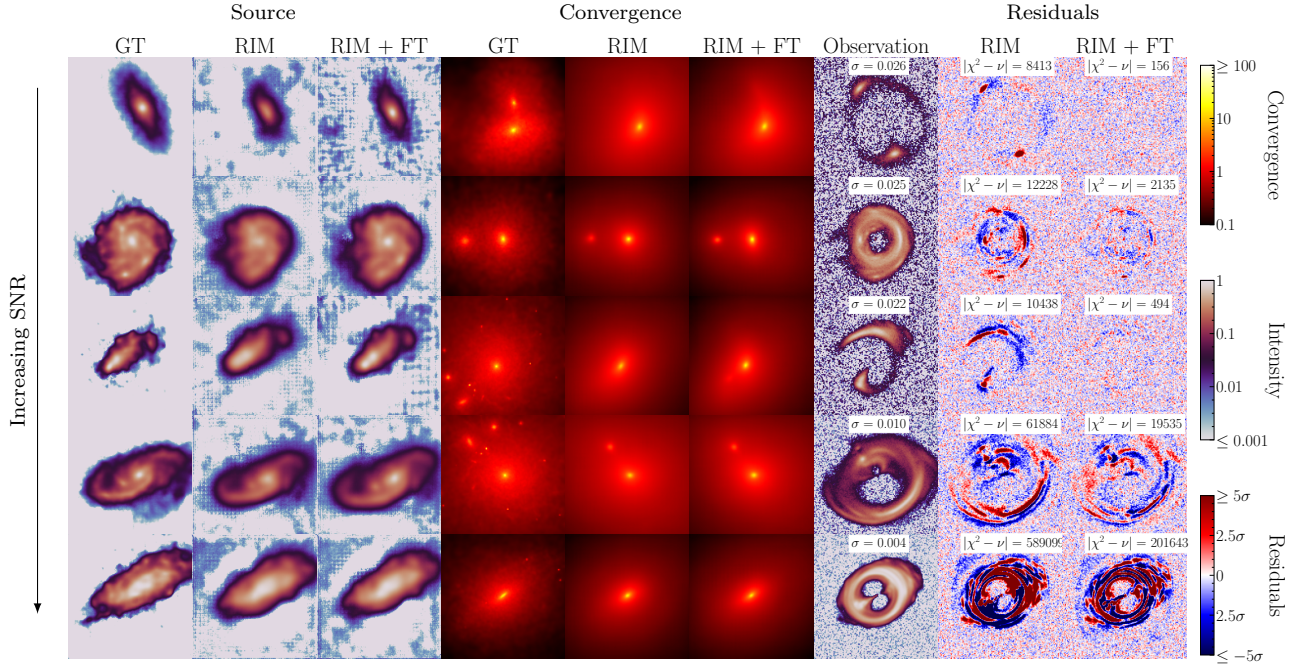

The right panel of Figure \ref{fig:loss and chi squared} shows the distribution of  $\chi^2$ for the test set before and after the fine-tuning procedure and the theoretical $\chi^2$ distribution corresponding to $\nu=128^2$ degrees of freedom.
We observe that the fine-tuning procedure significantly improves our $\chi^2$, bringing their distribution closer to that of the expected $\chi^2$ distribution (black curve). However, the improved distribution is still far from the theoretical expectation, implying that there are statistically significant residuals in a subset of the reconstructions.

In figure \ref{fig:chi squared vs noise}, we explore how the goodness of fit of the fine-tuned RIM changes as a function of SNR over the examples in the test set. Two behaviors can be identified. For SNR below a certain threshold, the goodness of fit 
of the fine-tuned model is essentially flat, with a certain scatter, around the noise level. This scatter increases as a function of SNR, which reflects the fact that above a certain SNR threshold (vertical dashed line in Figure \ref{fig:chi squared vs noise}), our reconstructions are dominated by systematics in the inference algorithm.
For SNR above the threshold, 
the goodness of fit follows the trend $\chi^2 \propto \sigma^{-2}$ (the solid line in Figure \ref{fig:chi squared vs noise}), which 
means the reconstructions have stopped improving on par with the SNR.

This behavior is exhibited in a few examples of reconstructions taken from the test set in Figure \ref{fig:increasing SNR}, where we order reconstructions with increasing SNR from top to bottom and plot the surface brightness and foreground densities in log scale. As can been seen, 
the amplitude of the residual increases significantly as we increase the SNR. Above the SNR threshold ($\sim 220$), the reconstructions are dominated by systematics.

\subsection{Quality of the Reconstructions}\label{sec:quality of reconstructions}

\begin{figure}[t!]
        \centering
        \includegraphics[width=\linewidth]{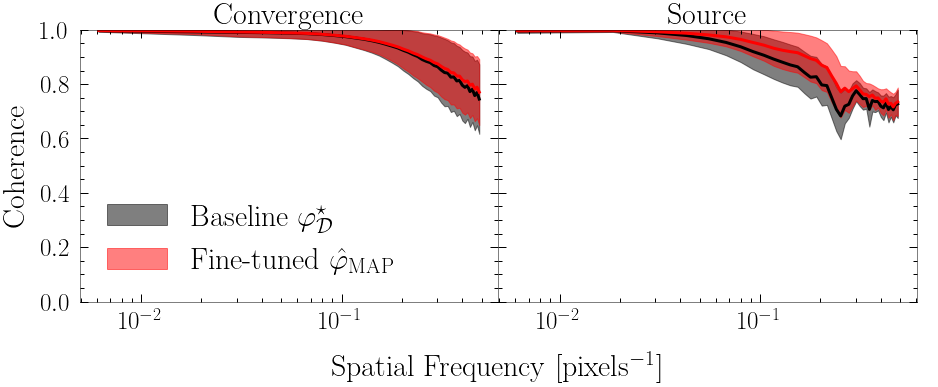}
        \caption{Statistics of the coherence spectrum on the test set. The solid line is the average 
        coherence. The transparent region is the $68\%$ confidence interval. The fine-tuning 
        procedure yields a noticeable improvement on the coherence of the source at all frequencies.}
        \label{fig:coherence}
\end{figure}

In addition to a visual inspection of the reconstructed sources 
and convergences, we compute 
the coherence spectrum to quantitatively assess the quality of the reconstructions
\begin{equation}\label{eq:coherence} 
        \gamma(k) = \frac{P_{12}(k)}{\sqrt{P_{11}(k) P_{22}(k)}} \, .
\end{equation}

Here, $P_{12}(k)$ is the cross power spectrum between a reconstructed and a true image at 
the wavenumber $k$. Figure \ref{fig:coherence} shows the coherence of the source and convergence maps 
for the entire test set of 3000 examples, 
with the mean value and the $68\%$ inclusion interval of $\gamma(k)$ reported in the solid line and shaded area respectively.
The fine-tuning procedure, shown in red, is able to significantly improve the coherence of the baseline background 
source, shown in black, at all scales. 
The coherence spectrum of the convergence sees a slight improvement due to the fine-tuning procedure.
Still, we note that many examples in the dataset exhibit significant 
improvement, which we illustrate in Figure \ref{fig:main figure}.
\pagebreak

\section{Conclusion}\label{sec:conclusion}
The results obtained here demonstrate the effectiveness of machine learning methods, specifically a recurrent inference machine, for inferring pixelated maps of the distribution of mass in lensing galaxies and the distribution of surface brightness in the background galaxies. Since this is a heavily under-constrained problem, stringent priors are needed to avoid overfitting the data, a task that has traditionally been difficult to accomplish with traditional statistical models \citep[e.g., ][]{Saha1997}. The model proposed here can implicitly learn these priors from a set of training data. 

The fine-tuning step that we propose in this work is a general procedure (i.e.\ not specific to our model or problem), which enables us to exploit a diagonal second-order Laplace approximation of the implicit prior learned by a baseline estimator during pre-training. We use fine-tuning in order to significantly improve this baseline estimator (i.e., a better MAP estimate), by using the likelihood of the data and the EWC prior. In the context of our work, we find that fine-tuning has a limiting --- or threshold --- behavior, which we speculate is due to the limited expressivity of the neural network and its inductive biases learned during pre-training.

The flexible and expressive form of the reconstructions shown in this work means that, in principle, any lensing system (e.g., a single simple galaxy or a group of complex galaxies) could be analyzed by this model, without any need for pre-determining the model parameterization. This is of high value given the diversity of observed lensing systems, and their relevance for constraining astrophysical and cosmological parameters. 

Perhaps the most important limitation of the method is the fact that, in its current form, the model only provides point estimates of the parameters of interest. Quantifying the posteriors of such high-dimensional data will require an efficient and accurate generative process \citep[e.g., see ][]{Adam:22a}, which we plan to explore and develop in future works.

\section*{Software and data}
The source code, as well as the various scripts and parameters used to 
produce the model and results is available as open-source software 
under the package \texttt{Censai}\footnote{
\href{https://github.com/AlexandreAdam/Censai}{
\includegraphics[scale=0.25]{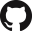}
https://github.com/AlexandreAdam/Censai}}. 
The model parameters, as well as convergence maps used to train 
these models and the test set examples and reconstructions results are also available as open-source datasets hosted by Zenodo\footnote{\href{https://doi.org/10.5281/zenodo.6555463}
{\includegraphics[scale=0.1]{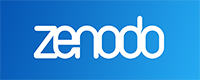}
https://doi.org/10.5281/zenodo.6555463}}. This research made use of \texttt{Tensorflow} \citep{tensorflow}, 
\texttt{Tensorflow-Probability} \citep{tensorflow-probability}, 
\texttt{Numpy} \citep{numpy}, 
\texttt{Scipy} \citep{scipy}, 
\texttt{Matplotlib} \citep{matplotlib}, 
\texttt{Scikit-image} \citep{scikit-image}, 
\texttt{IPython} \citep{ipython}, 
\texttt{Pandas} \citep{pandas1,pandas2}, 
\texttt{Scikit-learn} \citep{scikit-learn}, 
\texttt{Astropy} \citep{astropy:2013,astropy:2018} 
and \texttt{GalSim} \citep{galsim}.

\section*{Acknowledgements}
This research was made possible by a generous donation by Eric and Wendy Schmidt with the recommendation of the Schmidt Futures Foundation.

We would like to thank Ronan Legin for fruitful discussions and insights about training the neural network. The work is in part supported by computational resources provided by Calcul Quebec, Compute Canada and the Digital Research Alliance of Canada. Y.H. and L.P. acknowledge support from the National Sciences and Engineering Council of Canada grant RGPIN-2020-05102, the Fonds de recherche du Québec grant 2022-NC-301305 and 300397, and the Canada Research Chairs Program. A.A. was supported by an IVADO Excellence Scholarship.
\clearpage

\bibliography{bibliography}

\begin{thebibliography}{}
\expandafter\ifx\csname natexlab\endcsname\relax\def\natexlab#1{#1}\fi
\providecommand{\url}[1]{\href{#1}{#1}}
\providecommand{\dodoi}[1]{doi:~\href{http://doi.org/#1}{\nolinkurl{#1}}}
\providecommand{\doeprint}[1]{\href{http://ascl.net/#1}{\nolinkurl{http://ascl.net/#1}}}
\providecommand{\doarXiv}[1]{\href{https://arxiv.org/abs/#1}{\nolinkurl{https://arxiv.org/abs/#1}}}

\bibitem[{Abadi {et~al.}(2015)Abadi, Agarwal, Barham, Brevdo, Chen, Citro,
  Corrado, Davis, Dean, Devin, Ghemawat, Goodfellow, Harp, Irving, Isard, Jia,
  Jozefowicz, Kaiser, Kudlur, Levenberg, Man\'{e}, Monga, Moore, Murray, Olah,
  Schuster, Shlens, Steiner, Sutskever, Talwar, Tucker, Vanhoucke, Vasudevan,
  Vi\'{e}gas, Vinyals, Warden, Wattenberg, Wicke, Yu, \& Zheng}]{tensorflow}
Abadi, M., Agarwal, A., Barham, P., {et~al.} 2015, {TensorFlow}: Large-Scale
  Machine Learning on Heterogeneous Systems.
\newblock \url{https://www.tensorflow.org/}

\bibitem[{{Abdelsalam} {et~al.}(1998{\natexlab{a}}){Abdelsalam}, {Saha}, \&
  {Williams}}]{Abdelsalam1998}
{Abdelsalam}, H.~M., {Saha}, P., \& {Williams}, L. L.~R. 1998{\natexlab{a}},
  \aj, 116, 1541, \dodoi{10.1086/300546}

\bibitem[{{Abdelsalam} {et~al.}(1998{\natexlab{b}}){Abdelsalam}, {Saha}, \&
  {Williams}}]{Abdelsalam1998b}
---. 1998{\natexlab{b}}, \mnras, 294, 734,
  \dodoi{10.1046/j.1365-8711.1998.01356.x}

\bibitem[{{Adam} {et~al.}(2022){Adam}, {Coogan}, {Malkin}, {Legin},
  {Perreault-Levasseur}, {Hezaveh}, \& {Bengio}}]{Adam:22a}
{Adam}, A., {Coogan}, A., {Malkin}, N., {et~al.} 2022, arXiv e-prints,
  arXiv:2211.03812.
\newblock \doarXiv{2211.03812}

\bibitem[{{Anau Montel} {et~al.}(2022){Anau Montel}, {Coogan}, {Correa},
  {Karchev}, \& {Weniger}}]{AnauMontel2022}
{Anau Montel}, N., {Coogan}, A., {Correa}, C., {Karchev}, K., \& {Weniger}, C.
  2022, arXiv e-prints, arXiv:2205.09126.
\newblock \doarXiv{2205.09126}

\bibitem[{{Andrychowicz} {et~al.}(2016){Andrychowicz}, {Denil}, {Gomez},
  {Hoffman}, {Pfau}, {Schaul}, {Shillingford}, \& {de
  Freitas}}]{Andrychowicz2016}
{Andrychowicz}, M., {Denil}, M., {Gomez}, S., {et~al.} 2016, arXiv e-prints,
  arXiv:1606.04474.
\newblock \doarXiv{1606.04474}

\bibitem[{{Astropy Collaboration} {et~al.}(2013){Astropy Collaboration},
  {Robitaille}, {Tollerud}, {Greenfield}, {Droettboom}, {Bray}, {Aldcroft},
  {Davis}, {Ginsburg}, {Price-Whelan}, {Kerzendorf}, {Conley}, {Crighton},
  {Barbary}, {Muna}, {Ferguson}, {Grollier}, {Parikh}, {Nair}, {Unther},
  {Deil}, {Woillez}, {Conseil}, {Kramer}, {Turner}, {Singer}, {Fox}, {Weaver},
  {Zabalza}, {Edwards}, {Azalee Bostroem}, {Burke}, {Casey}, {Crawford},
  {Dencheva}, {Ely}, {Jenness}, {Labrie}, {Lim}, {Pierfederici}, {Pontzen},
  {Ptak}, {Refsdal}, {Servillat}, \& {Streicher}}]{astropy:2013}
{Astropy Collaboration}, {Robitaille}, T.~P., {Tollerud}, E.~J., {et~al.} 2013,
  \aap, 558, A33, \dodoi{10.1051/0004-6361/201322068}

\bibitem[{{Astropy Collaboration} {et~al.}(2018){Astropy Collaboration},
  {Price-Whelan}, {Sip{\H{o}}cz}, {G{\"u}nther}, {Lim}, {Crawford}, {Conseil},
  {Shupe}, {Craig}, {Dencheva}, {Ginsburg}, {Vand erPlas}, {Bradley},
  {P{\'e}rez-Su{\'a}rez}, {de Val-Borro}, {Aldcroft}, {Cruz}, {Robitaille},
  {Tollerud}, {Ardelean}, {Babej}, {Bach}, {Bachetti}, {Bakanov}, {Bamford},
  {Barentsen}, {Barmby}, {Baumbach}, {Berry}, {Biscani}, {Boquien}, {Bostroem},
  {Bouma}, {Brammer}, {Bray}, {Breytenbach}, {Buddelmeijer}, {Burke},
  {Calderone}, {Cano Rodr{\'\i}guez}, {Cara}, {Cardoso}, {Cheedella}, {Copin},
  {Corrales}, {Crichton}, {D'Avella}, {Deil}, {Depagne}, {Dietrich}, {Donath},
  {Droettboom}, {Earl}, {Erben}, {Fabbro}, {Ferreira}, {Finethy}, {Fox},
  {Garrison}, {Gibbons}, {Goldstein}, {Gommers}, {Greco}, {Greenfield},
  {Groener}, {Grollier}, {Hagen}, {Hirst}, {Homeier}, {Horton}, {Hosseinzadeh},
  {Hu}, {Hunkeler}, {Ivezi{\'c}}, {Jain}, {Jenness}, {Kanarek}, {Kendrew},
  {Kern}, {Kerzendorf}, {Khvalko}, {King}, {Kirkby}, {Kulkarni}, {Kumar},
  {Lee}, {Lenz}, {Littlefair}, {Ma}, {Macleod}, {Mastropietro}, {McCully},
  {Montagnac}, {Morris}, {Mueller}, {Mumford}, {Muna}, {Murphy}, {Nelson},
  {Nguyen}, {Ninan}, {N{\"o}the}, {Ogaz}, {Oh}, {Parejko}, {Parley}, {Pascual},
  {Patil}, {Patil}, {Plunkett}, {Prochaska}, {Rastogi}, {Reddy Janga},
  {Sabater}, {Sakurikar}, {Seifert}, {Sherbert}, {Sherwood-Taylor}, {Shih},
  {Sick}, {Silbiger}, {Singanamalla}, {Singer}, {Sladen}, {Sooley},
  {Sornarajah}, {Streicher}, {Teuben}, {Thomas}, {Tremblay}, {Turner},
  {Terr{\'o}n}, {van Kerkwijk}, {de la Vega}, {Watkins}, {Weaver}, {Whitmore},
  {Woillez}, {Zabalza}, \& {Astropy Contributors}}]{astropy:2018}
{Astropy Collaboration}, {Price-Whelan}, A.~M., {Sip{\H{o}}cz}, B.~M., {et~al.}
  2018, \aj, 156, 123, \dodoi{10.3847/1538-3881/aabc4f}

\bibitem[{Aubert {et~al.}(2007)Aubert, Amara, \& {Benton Metcalf}}]{Aubert2007}
Aubert, D., Amara, A., \& {Benton Metcalf}, R. 2007, Monthly Notices of the
  Royal Astronomical Society, 376, 113,
  \dodoi{10.1111/j.1365-2966.2006.11296.x}

\bibitem[{{Auger} {et~al.}(2010){Auger}, {Treu}, {Bolton}, {Gavazzi},
  {Koopmans}, {Marshall}, {Moustakas}, \& {Burles}}]{Auger2010}
{Auger}, M.~W., {Treu}, T., {Bolton}, A.~S., {et~al.} 2010, \apj, 724, 511,
  \dodoi{10.1088/0004-637X/724/1/511}

\bibitem[{{Barnab{\`e}} {et~al.}(2009){Barnab{\`e}}, {Czoske}, {Koopmans},
  {Treu}, {Bolton}, \& {Gavazzi}}]{Barnabe2009}
{Barnab{\`e}}, M., {Czoske}, O., {Koopmans}, L. V.~E., {et~al.} 2009, \mnras,
  399, 21, \dodoi{10.1111/j.1365-2966.2009.14941.x}

\bibitem[{{Bartelmann} {et~al.}(1996){Bartelmann}, {Narayan}, {Seitz}, \&
  {Schneider}}]{Bartelmann1996}
{Bartelmann}, M., {Narayan}, R., {Seitz}, S., \& {Schneider}, P. 1996, \apjl,
  464, L115, \dodoi{10.1086/310114}

\bibitem[{Bellagamba {et~al.}(2016)Bellagamba, Tessore, \&
  Metcalf}]{Bellagamba2016}
Bellagamba, F., Tessore, N., \& Metcalf, R.~B. 2016, Monthly Notices of the
  Royal Astronomical Society, 464, 4823, \dodoi{10.1093/mnras/stw2726}

\bibitem[{{Belokurov} {et~al.}(2007){Belokurov}, {Evans}, {Moiseev}, {King},
  {Hewett}, {Pettini}, {Wyrzykowski}, {McMahon}, {Smith}, {Gilmore}, {Sanchez},
  {Udalski}, {Koposov}, {Zucker}, \& {Walcher}}]{Belokurov2007}
{Belokurov}, V., {Evans}, N.~W., {Moiseev}, A., {et~al.} 2007, \apjl, 671, L9,
  \dodoi{10.1086/524948}

\bibitem[{Bengio(2009)}]{Bengio2009}
Bengio, Y. 2009, Found. Trends Mach. Learn., 2, 1–127,
  \dodoi{10.1561/2200000006}

\bibitem[{{Birrer} \& {Amara}(2018)}]{Birrer2018}
{Birrer}, S., \& {Amara}, A. 2018, Physics of the Dark Universe, 22, 189,
  \dodoi{10.1016/j.dark.2018.11.002}

\bibitem[{{Birrer} {et~al.}(2015){Birrer}, {Amara}, \&
  {Refregier}}]{Birrer2015}
{Birrer}, S., {Amara}, A., \& {Refregier}, A. 2015, \apj, 813, 102,
  \dodoi{10.1088/0004-637X/813/2/102}

\bibitem[{{Birrer} {et~al.}(2019){Birrer}, {Treu}, {Rusu}, {Bonvin},
  {Fassnacht}, {Chan}, {Agnello}, {Shajib}, {Chen}, {Auger}, {Courbin},
  {Hilbert}, {Sluse}, {Suyu}, {Wong}, {Marshall}, {Lemaux}, \&
  {Meylan}}]{Birrer2019}
{Birrer}, S., {Treu}, T., {Rusu}, C.~E., {et~al.} 2019, \mnras, 484, 4726,
  \dodoi{10.1093/mnras/stz200}

\bibitem[{{Bowman} {et~al.}(2015){Bowman}, {Vilnis}, {Vinyals}, {Dai},
  {Jozefowicz}, \& {Bengio}}]{Bowman2015}
{Bowman}, S.~R., {Vilnis}, L., {Vinyals}, O., {et~al.} 2015, arXiv e-prints,
  arXiv:1511.06349.
\newblock \doarXiv{1511.06349}

\bibitem[{{Brada{\v{c}}} {et~al.}(2005){Brada{\v{c}}}, {Schneider}, {Lombardi},
  \& {Erben}}]{Bradac2005}
{Brada{\v{c}}}, M., {Schneider}, P., {Lombardi}, M., \& {Erben}, T. 2005, \aap,
  437, 39, \dodoi{10.1051/0004-6361:20042233}

\bibitem[{{Burgess} {et~al.}(2018){Burgess}, {Higgins}, {Pal}, {Matthey},
  {Watters}, {Desjardins}, \& {Lerchner}}]{Burgess2018}
{Burgess}, C.~P., {Higgins}, I., {Pal}, A., {et~al.} 2018, arXiv e-prints,
  arXiv:1804.03599.
\newblock \doarXiv{1804.03599}

\bibitem[{{Cacciato} {et~al.}(2006){Cacciato}, {Bartelmann}, {Meneghetti}, \&
  {Moscardini}}]{Cacciato2006}
{Cacciato}, M., {Bartelmann}, M., {Meneghetti}, M., \& {Moscardini}, L. 2006,
  \aap, 458, 349, \dodoi{10.1051/0004-6361:20054582}

\bibitem[{Casella \& Berger(2001)}]{Casella2001}
Casella, G., \& Berger, R. 2001, Statistical Inference ({Duxbury Resource
  Center})

\bibitem[{{Cheng} {et~al.}(2019){Cheng}, {Wiesner}, {Peng}, {Cui}, {Peterson},
  \& {Li}}]{Cheng2019}
{Cheng}, J., {Wiesner}, M.~P., {Peng}, E.-H., {et~al.} 2019, \apj, 872, 185,
  \dodoi{10.3847/1538-4357/ab0029}

\bibitem[{{Cho} {et~al.}(2014){Cho}, {van Merrienboer}, {Gulcehre}, {Bahdanau},
  {Bougares}, {Schwenk}, \& {Bengio}}]{Cho2014}
{Cho}, K., {van Merrienboer}, B., {Gulcehre}, C., {et~al.} 2014, arXiv
  e-prints, arXiv:1406.1078.
\newblock \doarXiv{1406.1078}

\bibitem[{{Coe} {et~al.}(2008){Coe}, {Fuselier}, {Ben{\'\i}tez}, {Broadhurst},
  {Frye}, \& {Ford}}]{Coe2008}
{Coe}, D., {Fuselier}, E., {Ben{\'\i}tez}, N., {et~al.} 2008, \apj, 681, 814,
  \dodoi{10.1086/588250}

\bibitem[{{Coles} {et~al.}(2014){Coles}, {Read}, \& {Saha}}]{Coles2014}
{Coles}, J.~P., {Read}, J.~I., \& {Saha}, P. 2014, \mnras, 445, 2181,
  \dodoi{10.1093/mnras/stu1781}

\bibitem[{{Coogan} {et~al.}(2020){Coogan}, {Karchev}, \&
  {Weniger}}]{Coogan2020}
{Coogan}, A., {Karchev}, K., \& {Weniger}, C. 2020, arXiv e-prints,
  arXiv:2010.07032.
\newblock \doarXiv{2010.07032}

\bibitem[{{Cram\'{e}r}(1946)}]{Cramer1946}
{Cram\'{e}r}, H. 1946, {Mathematical methods of statistics}, Vol.~9 (Princeton
  University Press, Princeton, NJ)

\bibitem[{{Dalal} \& {Kochanek}(2002)}]{Dala2002}
{Dalal}, N., \& {Kochanek}, C.~S. 2002, \apj, 572, 25, \dodoi{10.1086/340303}

\bibitem[{{Deb} {et~al.}(2012){Deb}, {Morandi}, {Pedersen}, {Riemer-Sorensen},
  {Goldberg}, \& {Dahle}}]{Deb2012}
{Deb}, S., {Morandi}, A., {Pedersen}, K., {et~al.} 2012, arXiv e-prints,
  arXiv:1201.3636.
\newblock \doarXiv{1201.3636}

\bibitem[{{Diego} {et~al.}(2005){Diego}, {Protopapas}, {Sandvik}, \&
  {Tegmark}}]{Diego2005}
{Diego}, J.~M., {Protopapas}, P., {Sandvik}, H.~B., \& {Tegmark}, M. 2005,
  \mnras, 360, 477, \dodoi{10.1111/j.1365-2966.2005.09021.x}

\bibitem[{{Diego} {et~al.}(2007){Diego}, {Tegmark}, {Protopapas}, \&
  {Sandvik}}]{Diego2007}
{Diego}, J.~M., {Tegmark}, M., {Protopapas}, P., \& {Sandvik}, H.~B. 2007,
  \mnras, 375, 958, \dodoi{10.1111/j.1365-2966.2007.11380.x}

\bibitem[{{Dillon} {et~al.}(2017){Dillon}, {Langmore}, {Tran}, {Brevdo},
  {Vasudevan}, {Moore}, {Patton}, {Alemi}, {Hoffman}, \&
  {Saurous}}]{tensorflow-probability}
{Dillon}, J.~V., {Langmore}, I., {Tran}, D., {et~al.} 2017, arXiv e-prints,
  arXiv:1711.10604.
\newblock \doarXiv{1711.10604}

\bibitem[{{Galan} {et~al.}(2021){Galan}, {Peel}, {Joseph}, {Courbin}, \&
  {Starck}}]{Galan2021}
{Galan}, A., {Peel}, A., {Joseph}, R., {Courbin}, F., \& {Starck}, J.~L. 2021,
  \aap, 647, A176, \dodoi{10.1051/0004-6361/202039363}

\bibitem[{{Galan} {et~al.}(2022){Galan}, {Vernardos}, {Peel}, {Courbin}, \&
  {Starck}}]{Galan2022}
{Galan}, A., {Vernardos}, G., {Peel}, A., {Courbin}, F., \& {Starck}, J.~L.
  2022, \aap, 668, A155, \dodoi{10.1051/0004-6361/202244464}

\bibitem[{{Ghosh} {et~al.}(2020){Ghosh}, {Williams}, \&
  {Liesenborgs}}]{Ghosh2020}
{Ghosh}, A., {Williams}, L. L.~R., \& {Liesenborgs}, J. 2020, \mnras, 494,
  3998, \dodoi{10.1093/mnras/staa962}

\bibitem[{{Gilman} {et~al.}(2020){Gilman}, {Birrer}, {Nierenberg}, {Treu},
  {Du}, \& {Benson}}]{Gilman2020}
{Gilman}, D., {Birrer}, S., {Nierenberg}, A., {et~al.} 2020, \mnras, 491, 6077,
  \dodoi{10.1093/mnras/stz3480}

\bibitem[{{Gilman} {et~al.}(2021){Gilman}, {Bovy}, {Treu}, {Nierenberg},
  {Birrer}, {Benson}, \& {Sameie}}]{Gilman2021}
{Gilman}, D., {Bovy}, J., {Treu}, T., {et~al.} 2021, \mnras, 507, 2432,
  \dodoi{10.1093/mnras/stab2335}

\bibitem[{Harris {et~al.}(2020)Harris, Millman, van~der Walt, Gommers,
  Virtanen, Cournapeau, Wieser, Taylor, Berg, Smith, Kern, Picus, Hoyer, van
  Kerkwijk, Brett, Haldane, del R{\'{i}}o, Wiebe, Peterson,
  G{\'{e}}rard-Marchant, Sheppard, Reddy, Weckesser, Abbasi, Gohlke, \&
  Oliphant}]{numpy}
Harris, C.~R., Millman, K.~J., van~der Walt, S.~J., {et~al.} 2020, Nature, 585,
  357, \dodoi{10.1038/s41586-020-2649-2}

\bibitem[{{Hezaveh} {et~al.}(2017){Hezaveh}, {Perreault Levasseur}, \&
  {Marshall}}]{Hezaveh2017}
{Hezaveh}, Y.~D., {Perreault Levasseur}, L., \& {Marshall}, P.~J. 2017, \nat,
  548, 555, \dodoi{10.1038/nature23463}

\bibitem[{{Hezaveh} {et~al.}(2016){Hezaveh}, {Dalal}, {Marrone}, {Mao},
  {Morningstar}, {Wen}, {Blandford}, {Carlstrom}, {Fassnacht}, {Holder},
  {Kemball}, {Marshall}, {Murray}, {Perreault Levasseur}, {Vieira}, \&
  {Wechsler}}]{Hezaveh2016}
{Hezaveh}, Y.~D., {Dalal}, N., {Marrone}, D.~P., {et~al.} 2016, \apj, 823, 37,
  \dodoi{10.3847/0004-637X/823/1/37}

\bibitem[{Higgins {et~al.}(2017)Higgins, Matthey, Pal, Burgess, Glorot,
  Botvinick, Mohamed, \& Lerchner}]{Higgins2017}
Higgins, I., Matthey, L., Pal, A., {et~al.} 2017, in ICLR

\bibitem[{Hunter(2007)}]{matplotlib}
Hunter, J.~D. 2007, Computing in Science \& Engineering, 9, 90,
  \dodoi{10.1109/MCSE.2007.55}

\bibitem[{{Jee} {et~al.}(2007){Jee}, {Ford}, {Illingworth}, {White},
  {Broadhurst}, {Coe}, {Meurer}, {van der Wel}, {Ben{\'\i}tez}, {Blakeslee},
  {Bouwens}, {Bradley}, {Demarco}, {Homeier}, {Martel}, \& {Mei}}]{Jee2007}
{Jee}, M.~J., {Ford}, H.~C., {Illingworth}, G.~D., {et~al.} 2007, \apj, 661,
  728, \dodoi{10.1086/517498}

\bibitem[{{Kaae S{\o}nderby} {et~al.}(2016){Kaae S{\o}nderby}, {Raiko},
  {Maal{\o}e}, {Kaae S{\o}nderby}, \& {Winther}}]{Sonderby2016}
{Kaae S{\o}nderby}, C., {Raiko}, T., {Maal{\o}e}, L., {Kaae S{\o}nderby}, S.,
  \& {Winther}, O. 2016, arXiv e-prints, arXiv:1602.02282.
\newblock \doarXiv{1602.02282}

\bibitem[{{Karchev} {et~al.}(2022){Karchev}, {Coogan}, \&
  {Weniger}}]{Karchev2022}
{Karchev}, K., {Coogan}, A., \& {Weniger}, C. 2022, \mnras, 512, 661,
  \dodoi{10.1093/mnras/stac311}

\bibitem[{{Kingma} \& {Ba}(2014)}]{Kingma2014}
{Kingma}, D.~P., \& {Ba}, J. 2014, arXiv e-prints, arXiv:1412.6980.
\newblock \doarXiv{1412.6980}

\bibitem[{{Kingma} \& {Welling}(2013)}]{Kingma2013}
{Kingma}, D.~P., \& {Welling}, M. 2013, arXiv e-prints, arXiv:1312.6114.
\newblock \doarXiv{1312.6114}

\bibitem[{{Kingma} \& {Welling}(2019)}]{Kingma2019}
---. 2019, arXiv e-prints, arXiv:1906.02691.
\newblock \doarXiv{1906.02691}

\bibitem[{{Kirkpatrick} {et~al.}(2016){Kirkpatrick}, {Pascanu}, {Rabinowitz},
  {Veness}, {Desjardins}, {Rusu}, {Milan}, {Quan}, {Ramalho},
  {Grabska-Barwinska}, {Hassabis}, {Clopath}, {Kumaran}, \&
  {Hadsell}}]{Kirkpatrick2016}
{Kirkpatrick}, J., {Pascanu}, R., {Rabinowitz}, N., {et~al.} 2016, arXiv
  e-prints, arXiv:1612.00796.
\newblock \doarXiv{1612.00796}

\bibitem[{Koekemoer {et~al.}(2007)Koekemoer, Aussel, Calzetti, Capak,
  Giavalisco, Kneib, Leauthaud, {Le Fevre}, McCracken, Massey, Mobasher,
  Rhodes, Scoville, \& Shopbell}]{Koekemoer2007}
Koekemoer, A.~M., Aussel, H., Calzetti, D., {et~al.} 2007, The Astrophysical
  Journal Supplement Series, 172, 196, \dodoi{10.1086/520086}

\bibitem[{{Koopmans} {et~al.}(2006){Koopmans}, {Treu}, {Bolton}, {Burles}, \&
  {Moustakas}}]{Koopmans2006}
{Koopmans}, L. V.~E., {Treu}, T., {Bolton}, A.~S., {Burles}, S., \&
  {Moustakas}, L.~A. 2006, \apj, 649, 599, \dodoi{10.1086/505696}

\bibitem[{{Lanusse} {et~al.}(2021){Lanusse}, {Mandelbaum}, {Ravanbakhsh}, {Li},
  {Freeman}, \& {P{\'o}czos}}]{Lanusse2021}
{Lanusse}, F., {Mandelbaum}, R., {Ravanbakhsh}, S., {et~al.} 2021, \mnras, 504,
  5543, \dodoi{10.1093/mnras/stab1214}

\bibitem[{Leauthaud {et~al.}(2007)Leauthaud, Massey, Kneib, Rhodes, Johnston,
  Capak, Heymans, Ellis, Koekemoer, F{\`{e}}vre, Mellier,
  R{\'{e}}fr{\'{e}}gier, Robin, Scoville, Tasca, Taylor, \&
  Waerbeke}]{Leauthaud2007}
Leauthaud, A., Massey, R., Kneib, J.-P., {et~al.} 2007, The Astrophysical
  Journal Supplement Series, 172, 219, \dodoi{10.1086/516598}

\bibitem[{{Legin} {et~al.}(2021){Legin}, {Hezaveh}, {Perreault Levasseur}, \&
  {Wandelt}}]{Legin2021}
{Legin}, R., {Hezaveh}, Y., {Perreault Levasseur}, L., \& {Wandelt}, B. 2021,
  arXiv e-prints, arXiv:2112.05278.
\newblock \doarXiv{2112.05278}

\bibitem[{{Legin} {et~al.}(2022){Legin}, {Stone}, {Hezaveh}, \&
  {Perreault-Levasseur}}]{Legin2022}
{Legin}, R., {Stone}, C., {Hezaveh}, Y., \& {Perreault-Levasseur}, L. 2022,
  arXiv e-prints, arXiv:2207.04123.
\newblock \doarXiv{2207.04123}

\bibitem[{{Li} {et~al.}(2021){Li}, {Becker}, \& {Dye}}]{Li2021}
{Li}, N., {Becker}, C., \& {Dye}, S. 2021, \mnras, 504, 2224,
  \dodoi{10.1093/mnras/stab984}

\bibitem[{{Liesenborgs} {et~al.}(2006){Liesenborgs}, {De Rijcke}, \&
  {Dejonghe}}]{Liesenborgs2006}
{Liesenborgs}, J., {De Rijcke}, S., \& {Dejonghe}, H. 2006, \mnras, 367, 1209,
  \dodoi{10.1111/j.1365-2966.2006.10040.x}

\bibitem[{{Liesenborgs} {et~al.}(2007){Liesenborgs}, {de Rijcke}, {Dejonghe},
  \& {Bekaert}}]{Liesenborgs2007}
{Liesenborgs}, J., {de Rijcke}, S., {Dejonghe}, H., \& {Bekaert}, P. 2007,
  \mnras, 380, 1729, \dodoi{10.1111/j.1365-2966.2007.12236.x}

\bibitem[{L{\o}nning {et~al.}(2019)L{\o}nning, Putzky, Sonke, Reneman, Caan, \&
  Welling}]{Lonning2019}
L{\o}nning, K., Putzky, P., Sonke, J.~J., {et~al.} 2019, Medical Image
  Analysis, 53, 64, \dodoi{10.1016/j.media.2019.01.005}

\bibitem[{Mandelbaum {et~al.}(2012)Mandelbaum, Lackner, Leauthaud, \&
  Rowe}]{Mandelbaum2012}
Mandelbaum, R., Lackner, C., Leauthaud, A., \& Rowe, B. 2012, Zenodo.
\newblock \url{https://zenodo.org/record/3242143}

\bibitem[{Mandelbaum {et~al.}(2014)Mandelbaum, Rowe, Bosch, Chang, Courbin,
  Gill, Jarvis, Kannawadi, Kacprzak, Lackner, Leauthaud, Miyatake, Nakajima,
  Rhodes, Simet, Zuntz, Armstrong, Bridle, Coupon, Dietrich, Gentile, Heymans,
  Jurling, Kent, Kirkby, Margala, Massey, Melchior, Peterson, Roodman, \&
  Schrabback}]{Mandelbaum2014}
Mandelbaum, R., Rowe, B., Bosch, J., {et~al.} 2014, The Astrophysical Journal
  Supplement Series, 212, 5, \dodoi{10.1088/0067-0049/212/1/5}

\bibitem[{{Marrone} {et~al.}(2018){Marrone}, {Spilker}, {Hayward}, {Vieira},
  {Aravena}, {Ashby}, {Bayliss}, {B{\'e}thermin}, {Brodwin}, {Bothwell},
  {Carlstrom}, {Chapman}, {Chen}, {Crawford}, {Cunningham}, {De Breuck},
  {Fassnacht}, {Gonzalez}, {Greve}, {Hezaveh}, {Lacaille}, {Litke}, {Lower},
  {Ma}, {Malkan}, {Miller}, {Morningstar}, {Murphy}, {Narayanan}, {Phadke},
  {Rotermund}, {Sreevani}, {Stalder}, {Stark}, {Strandet}, {Tang}, \&
  {Wei{\ss}}}]{Marrone2018}
{Marrone}, D.~P., {Spilker}, J.~S., {Hayward}, C.~C., {et~al.} 2018, \nat, 553,
  51, \dodoi{10.1038/nature24629}

\bibitem[{McCloskey \& Cohen(1989)}]{McCloskey1989}
McCloskey, M., \& Cohen, N.~J. 1989in  (Academic Press), 109--165,
  \dodoi{10.1016/S0079-7421(08)60536-8}

\bibitem[{{Merten}(2016)}]{Merten2016}
{Merten}, J. 2016, \mnras, 461, 2328, \dodoi{10.1093/mnras/stw1413}

\bibitem[{{Merten} {et~al.}(2009){Merten}, {Cacciato}, {Meneghetti}, {Mignone},
  \& {Bartelmann}}]{Merten2009}
{Merten}, J., {Cacciato}, M., {Meneghetti}, M., {Mignone}, C., \& {Bartelmann},
  M. 2009, \aap, 500, 681, \dodoi{10.1051/0004-6361/200810372}

\bibitem[{Mishra-Sharma \& Yang(2022)}]{Mishra-Sharma2022}
Mishra-Sharma, S., \& Yang, G. 2022, Strong Lensing Source Reconstruction Using
  Continuous Neural Fields,  arXiv, \dodoi{10.48550/ARXIV.2206.14820}

\bibitem[{{Modi} {et~al.}(2021){Modi}, {Lanusse}, {Seljak}, {Spergel}, \&
  {Perreault-Levasseur}}]{Modi2021}
{Modi}, C., {Lanusse}, F., {Seljak}, U., {Spergel}, D.~N., \&
  {Perreault-Levasseur}, L. 2021, arXiv e-prints, arXiv:2104.12864.
\newblock \doarXiv{2104.12864}

\bibitem[{Morningstar {et~al.}(2018)Morningstar, Hezaveh, Levasseur, Blandford,
  Marshall, Putzky, \& Wechsler}]{Morningstar2018}
Morningstar, W.~R., Hezaveh, Y.~D., Levasseur, L.~P., {et~al.} 2018, arXiv
  e-prints.
\newblock \doarXiv{1808.00011v1}

\bibitem[{Morningstar {et~al.}(2019)Morningstar, Levasseur, Hezaveh, Blandford,
  Marshall, Putzky, Rueter, Wechsler, \& Welling}]{Morningstar2019}
Morningstar, W.~R., Levasseur, L.~P., Hezaveh, Y.~D., {et~al.} 2019, The
  Astrophysical Journal, 883, 14, \dodoi{10.3847/1538-4357/ab35d7}

\bibitem[{Nelson {et~al.}(2019)Nelson, Springel, Pillepich, Rodriguez-Gomez,
  Torrey, Genel, Vogelsberger, Pakmor, Marinacci, Weinberger, Kelley, Lovell,
  Diemer, \& Hernquist}]{Nelson2018}
Nelson, D., Springel, V., Pillepich, A., {et~al.} 2019, \mnras, 6,
  \dodoi{10.1186/s40668-019-0028-x}

\bibitem[{{Nightingale} {et~al.}(2018){Nightingale}, {Dye}, \&
  {Massey}}]{Nightingale2018}
{Nightingale}, J.~W., {Dye}, S., \& {Massey}, R.~J. 2018, \mnras, 478, 4738,
  \dodoi{10.1093/mnras/sty1264}

\bibitem[{Pan \& Yang(2010)}]{Pan2010}
Pan, S.~J., \& Yang, Q. 2010, IEEE Transactions on Knowledge and Data
  Engineering, 22, 1345

\bibitem[{pandas~development team(2020)}]{pandas2}
pandas~development team, T. 2020, pandas-dev/pandas: Pandas, latest,  Zenodo,
  \dodoi{10.5281/zenodo.3509134}

\bibitem[{{Park} {et~al.}(2021){Park}, {Wagner-Carena}, {Birrer}, {Marshall},
  {Lin}, {Roodman}, \& {LSST Dark Energy Science Collaboration}}]{Park2021}
{Park}, J.~W., {Wagner-Carena}, S., {Birrer}, S., {et~al.} 2021, \apj, 910, 39,
  \dodoi{10.3847/1538-4357/abdfc4}

\bibitem[{Pedregosa {et~al.}(2011)Pedregosa, Varoquaux, Gramfort, Michel,
  Thirion, Grisel, Blondel, Prettenhofer, Weiss, Dubourg, Vanderplas, Passos,
  Cournapeau, Brucher, Perrot, \& Duchesnay}]{scikit-learn}
Pedregosa, F., Varoquaux, G., Gramfort, A., {et~al.} 2011, Journal of Machine
  Learning Research, 12, 2825

\bibitem[{P\'erez \& Granger(2007)}]{ipython}
P\'erez, F., \& Granger, B.~E. 2007, Computing in Science and Engineering, 9,
  21, \dodoi{10.1109/MCSE.2007.53}

\bibitem[{{Perreault Levasseur} {et~al.}(2017){Perreault Levasseur}, {Hezaveh},
  \& {Wechsler}}]{PerreaultLevasseur2017}
{Perreault Levasseur}, L., {Hezaveh}, Y.~D., \& {Wechsler}, R.~H. 2017, \apjl,
  850, L7, \dodoi{10.3847/2041-8213/aa9704}

\bibitem[{{Planck Collaboration}(2020)}]{PlanckCollaboration2018}
{Planck Collaboration}. 2020, \aap, 641, A6,
  \dodoi{10.1051/0004-6361/201833910}

\bibitem[{Putzky \& Welling(2017)}]{Putzky2017}
Putzky, P., \& Welling, M. 2017, arXiv e-prints.
\newblock \doarXiv{1706.04008}

\bibitem[{{Rahaman} {et~al.}(2018){Rahaman}, {Baratin}, {Arpit}, {Draxler},
  {Lin}, {Hamprecht}, {Bengio}, \& {Courville}}]{Rahaman2018}
{Rahaman}, N., {Baratin}, A., {Arpit}, D., {et~al.} 2018, arXiv e-prints,
  arXiv:1806.08734.
\newblock \doarXiv{1806.08734}

\bibitem[{Rao(1945)}]{Rao1945}
Rao, C. 1945, Bulletin fo the Calcutta Mathematical Society

\bibitem[{Ratcliff(1990)}]{Ratcliff1990}
Ratcliff, R. 1990, Psychological Review, 97, 285,
  \dodoi{10.1037/0033-295X.97.2.285}

\bibitem[{Rau {et~al.}(2013)Rau, Vegetti, \& White}]{Rau2013}
Rau, S., Vegetti, S., \& White, S.~D. 2013, Monthly Notices of the Royal
  Astronomical Society, 430, 2232, \dodoi{10.1093/mnras/stt043}

\bibitem[{{Rizzo} {et~al.}(2020){Rizzo}, {Vegetti}, {Powell}, {Fraternali},
  {McKean}, {Stacey}, \& {White}}]{Rizzo2020}
{Rizzo}, F., {Vegetti}, S., {Powell}, D., {et~al.} 2020, \nat, 584, 201,
  \dodoi{10.1038/s41586-020-2572-6}

\bibitem[{{Ronneberger} {et~al.}(2015){Ronneberger}, {Fischer}, \&
  {Brox}}]{Ronneberger2015}
{Ronneberger}, O., {Fischer}, P., \& {Brox}, T. 2015, arXiv e-prints,
  arXiv:1505.04597.
\newblock \doarXiv{1505.04597}

\bibitem[{Rowe {et~al.}(2015)Rowe, Jarvis, Mandelbaum, Bernstein, Bosch, Simet,
  Meyers, Kacprzak, Nakajima, Zuntz, Miyatake, Dietrich, Armstrong, Melchior,
  \& Gill}]{Rowe2015}
Rowe, B.~T., Jarvis, M., Mandelbaum, R., {et~al.} 2015, Astronomy and
  Computing, 10, 121, \dodoi{10.1016/j.ascom.2015.02.002}

\bibitem[{{Rowe} {et~al.}(2015){Rowe}, {Jarvis}, {Mandelbaum}, {Bernstein},
  {Bosch}, {Simet}, {Meyers}, {Kacprzak}, {Nakajima}, {Zuntz}, {Miyatake},
  {Dietrich}, {Armstrong}, {Melchior}, \& {Gill}}]{galsim}
{Rowe}, B.~T.~P., {Jarvis}, M., {Mandelbaum}, R., {et~al.} 2015, Astronomy and
  Computing, 10, 121, \dodoi{10.1016/j.ascom.2015.02.002}

\bibitem[{{Rusu} {et~al.}(2017){Rusu}, {Fassnacht}, {Sluse}, {Hilbert}, {Wong},
  {Huang}, {Suyu}, {Collett}, {Marshall}, {Treu}, \& {Koopmans}}]{Rusu2017}
{Rusu}, C.~E., {Fassnacht}, C.~D., {Sluse}, D., {et~al.} 2017, \mnras, 467,
  4220, \dodoi{10.1093/mnras/stx285}

\bibitem[{{Rusu} {et~al.}(2020){Rusu}, {Wong}, {Bonvin}, {Sluse}, {Suyu},
  {Fassnacht}, {Chan}, {Hilbert}, {Auger}, {Sonnenfeld}, {Birrer}, {Courbin},
  {Treu}, {Chen}, {Halkola}, {Koopmans}, {Marshall}, \& {Shajib}}]{Rusu2019}
{Rusu}, C.~E., {Wong}, K.~C., {Bonvin}, V., {et~al.} 2020, \mnras, 498, 1440,
  \dodoi{10.1093/mnras/stz3451}

\bibitem[{{Saha} \& {Williams}(1997)}]{Saha1997}
{Saha}, P., \& {Williams}, L. L.~R. 1997, \mnras, 292, 148,
  \dodoi{10.1093/mnras/292.1.148}

\bibitem[{{Saha} \& {Williams}(2004)}]{Saha2004}
---. 2004, \aj, 127, 2604, \dodoi{10.1086/383544}

\bibitem[{{Schmidt} {et~al.}(2022){Schmidt}, {Treu}, {Birrer}, {Shajib},
  {Lemon}, {Millon}, {Sluse}, {Agnello}, {Anguita}, {Auger-Williams},
  {McMahon}, {Motta}, {Schechter}, {Spiniello}, {Kayo}, {Courbin}, {Ertl},
  {Fassnacht}, {Frieman}, {More}, {Schuldt}, {Suyu}, {Aguena},
  {Andrade-Oliveira}, {Annis}, {Bacon}, {Bertin}, {Brooks}, {Burke}, {Carnero
  Rosell}, {Carrasco Kind}, {Carretero}, {Conselice}, {Costanzi}, {da Costa},
  {Pereira}, {De Vicente}, {Desai}, {Doel}, {Everett}, {Ferrero}, {Friedel},
  {Garc{\'\i}a-Bellido}, {Gaztanaga}, {Gruen}, {Gruendl}, {Gschwend},
  {Gutierrez}, {Hinton}, {Hollowood}, {Honscheid}, {James}, {Kuehn}, {Lahav},
  {Menanteau}, {Miquel}, {Palmese}, {Paz-Chinch{\'o}n}, {Pieres}, {Plazas
  Malag{\'o}n}, {Prat}, {Rodriguez-Monroy}, {Romer}, {Sanchez}, {Scarpine},
  {Sevilla-Noarbe}, {Smith}, {Suchyta}, {Tarle}, {To}, \&
  {Varga}}]{Schmidt2022}
{Schmidt}, T., {Treu}, T., {Birrer}, S., {et~al.} 2022, arXiv e-prints,
  arXiv:2206.04696.
\newblock \doarXiv{2206.04696}

\bibitem[{{Schuldt} {et~al.}(2019){Schuldt}, {Chiriv{\`\i}}, {Suyu},
  {Y{\i}ld{\i}r{\i}m}, {Sonnenfeld}, {Halkola}, \& {Lewis}}]{Schuldt2019}
{Schuldt}, S., {Chiriv{\`\i}}, G., {Suyu}, S.~H., {et~al.} 2019, \aap, 631,
  A40, \dodoi{10.1051/0004-6361/201935042}

\bibitem[{{Schuldt} {et~al.}(2022){Schuldt}, {Suyu}, {Canameras}, {Shu},
  {Taubenberger}, {Ertl}, \& {Halkola}}]{Schuldt2022}
{Schuldt}, S., {Suyu}, S.~H., {Canameras}, R., {et~al.} 2022, arXiv e-prints,
  arXiv:2207.10124.
\newblock \doarXiv{2207.10124}

\bibitem[{Scoville {et~al.}(2007)Scoville, Aussel, Brusa, Capak, Carollo,
  Elvis, Giavalisco, Guzzo, Hasinger, Impey, Kneib, LeFevre, Lilly, Mobasher,
  Renzini, Rich, Sanders, Schinnerer, Schminovich, Shopbell, Taniguchi, \&
  Tyson}]{Scoville2007}
Scoville, N., Aussel, H., Brusa, M., {et~al.} 2007, The Astrophysical Journal
  Supplement Series, 172, 1, \dodoi{10.1086/516585}

\bibitem[{{Seitz} {et~al.}(1998){Seitz}, {Schneider}, \&
  {Bartelmann}}]{Seitz1998}
{Seitz}, S., {Schneider}, P., \& {Bartelmann}, M. 1998, \aap, 337, 325.
\newblock \doarXiv{astro-ph/9803038}

\bibitem[{{S{\'e}rsic}(1963)}]{Sersic1963}
{S{\'e}rsic}, J.~L. 1963, Boletin de la Asociacion Argentina de Astronomia La
  Plata Argentina, 6, 41

\bibitem[{{Sluse} {et~al.}(2017){Sluse}, {Sonnenfeld}, {Rumbaugh}, {Rusu},
  {Fassnacht}, {Treu}, {Suyu}, {Wong}, {Auger}, {Bonvin}, {Collett}, {Courbin},
  {Hilbert}, {Koopmans}, {Marshall}, {Meylan}, {Spiniello}, \&
  {Tewes}}]{Sluse2017}
{Sluse}, D., {Sonnenfeld}, A., {Rumbaugh}, N., {et~al.} 2017, \mnras, 470,
  4838, \dodoi{10.1093/mnras/stx1484}

\bibitem[{{Sun} {et~al.}(2021){Sun}, {Egami}, {P{\'e}rez-Gonz{\'a}lez},
  {Smail}, {Caputi}, {Bauer}, {Rawle}, {Fujimoto}, {Kohno},
  {Dudzevi{\v{c}}i{\={u}}t{\.{e}}}, {Atek}, {Bianconi}, {Chapman}, {Combes},
  {Jauzac}, {Jolly}, {Koekemoer}, {Magdis}, {Rodighiero}, {Rujopakarn},
  {Schaerer}, {Steinhardt}, {Van der Werf}, {Walth}, \& {Weaver}}]{Sun2021}
{Sun}, F., {Egami}, E., {P{\'e}rez-Gonz{\'a}lez}, P.~G., {et~al.} 2021, \apj,
  922, 114, \dodoi{10.3847/1538-4357/ac2578}

\bibitem[{{Suyu} {et~al.}(2006){Suyu}, {Marshall}, {Hobson}, \&
  {Blandford}}]{Suyu2006}
{Suyu}, S.~H., {Marshall}, P.~J., {Hobson}, M.~P., \& {Blandford}, R.~D. 2006,
  \mnras, 371, 983, \dodoi{10.1111/j.1365-2966.2006.10733.x}

\bibitem[{{Tishby} {et~al.}(2000){Tishby}, {Pereira}, \& {Bialek}}]{Tishby2000}
{Tishby}, N., {Pereira}, F.~C., \& {Bialek}, W. 2000, arXiv e-prints,
  physics/0004057.
\newblock \doarXiv{physics/0004057}

\bibitem[{{Torres-Ballesteros} \&
  {Casta{\~n}eda}(2022)}]{Torres-Ballestros2022}
{Torres-Ballesteros}, D.~A., \& {Casta{\~n}eda}, L. 2022, arXiv e-prints,
  arXiv:2201.10076.
\newblock \doarXiv{2201.10076}

\bibitem[{{Treu} \& {Koopmans}(2004)}]{Treu2004}
{Treu}, T., \& {Koopmans}, L. V.~E. 2004, \apj, 611, 739,
  \dodoi{10.1086/422245}

\bibitem[{Van~der Walt {et~al.}(2014)Van~der Walt, Sch{\"o}nberger,
  Nunez-Iglesias, Boulogne, Warner, Yager, Gouillart, \& Yu}]{scikit-image}
Van~der Walt, S., Sch{\"o}nberger, J.~L., Nunez-Iglesias, J., {et~al.} 2014,
  PeerJ, 2, e453

\bibitem[{{Vieira} {et~al.}(2013){Vieira}, {Marrone}, {Chapman}, {De Breuck},
  {Hezaveh}, {Wei{\ensuremath{\beta}}}, {Aguirre}, {Aird}, {Aravena}, {Ashby},
  {Bayliss}, {Benson}, {Biggs}, {Bleem}, {Bock}, {Bothwell}, {Bradford},
  {Brodwin}, {Carlstrom}, {Chang}, {Crawford}, {Crites}, {de Haan}, {Dobbs},
  {Fomalont}, {Fassnacht}, {George}, {Gladders}, {Gonzalez}, {Greve},
  {Gullberg}, {Halverson}, {High}, {Holder}, {Holzapfel}, {Hoover}, {Hrubes},
  {Hunter}, {Keisler}, {Lee}, {Leitch}, {Lueker}, {Luong-van}, {Malkan},
  {McIntyre}, {McMahon}, {Mehl}, {Menten}, {Meyer}, {Mocanu}, {Murphy},
  {Natoli}, {Padin}, {Plagge}, {Reichardt}, {Rest}, {Ruel}, {Ruhl}, {Sharon},
  {Schaffer}, {Shaw}, {Shirokoff}, {Spilker}, {Stalder}, {Staniszewski},
  {Stark}, {Story}, {Vanderlinde}, {Welikala}, \& {Williamson}}]{Viera2013}
{Vieira}, J.~D., {Marrone}, D.~P., {Chapman}, S.~C., {et~al.} 2013, \nat, 495,
  344, \dodoi{10.1038/nature12001}

\bibitem[{Vincent {et~al.}(2008)Vincent, Larochelle, Bengio, \&
  Manzagol}]{Vincent2008}
Vincent, P., Larochelle, H., Bengio, Y., \& Manzagol, P.-A. 2008, in
  Proceedings of the 25th International Conference on Machine Learning, ICML
  '08 (New York, NY, USA: Association for Computing Machinery), 1096–1103,
  \dodoi{10.1145/1390156.1390294}

\bibitem[{Vincent {et~al.}(2010)Vincent, Larochelle, Lajoie, Bengio, \&
  Manzagol}]{Vincent2010}
Vincent, P., Larochelle, H., Lajoie, I., Bengio, Y., \& Manzagol, P.-A. 2010,
  J. Mach. Learn. Res., 11, 3371–3408

\bibitem[{Virtanen {et~al.}(2020)Virtanen, Gommers, Oliphant, Haberland, Reddy,
  Cournapeau, Burovski, Peterson, Weckesser, Bright, {van der Walt}, Brett,
  Wilson, Millman, Mayorov, Nelson, Jones, Kern, Larson, Carey, Polat, Feng,
  Moore, {VanderPlas}, Laxalde, Perktold, Cimrman, Henriksen, Quintero, Harris,
  Archibald, Ribeiro, Pedregosa, {van Mulbregt}, \& {SciPy 1.0
  Contributors}}]{scipy}
Virtanen, P., Gommers, R., Oliphant, T.~E., {et~al.} 2020, Nature Methods, 17,
  261, \dodoi{10.1038/s41592-019-0686-2}

\bibitem[{{Wagner-Carena} {et~al.}(2022){Wagner-Carena}, {Aalbers}, {Birrer},
  {Nadler}, {Darragh-Ford}, {Marshall}, \& {Wechsler}}]{Wagner-Carena2022}
{Wagner-Carena}, S., {Aalbers}, J., {Birrer}, S., {et~al.} 2022, arXiv
  e-prints, arXiv:2203.00690.
\newblock \doarXiv{2203.00690}

\bibitem[{{Wagner-Carena} {et~al.}(2021){Wagner-Carena}, {Park}, {Birrer},
  {Marshall}, {Roodman}, {Wechsler}, \& {LSST Dark Energy Science
  Collaboration}}]{Wagner-Carena2021}
{Wagner-Carena}, S., {Park}, J.~W., {Birrer}, S., {et~al.} 2021, \apj, 909,
  187, \dodoi{10.3847/1538-4357/abdf59}

\bibitem[{{Warren} \& {Dye}(2003)}]{Warren2003}
{Warren}, S.~J., \& {Dye}, S. 2003, \apj, 590, 673, \dodoi{10.1086/375132}

\bibitem[{{W}es {M}c{K}inney(2010)}]{pandas1}
{W}es {M}c{K}inney. 2010, in {P}roceedings of the 9th {P}ython in {S}cience
  {C}onference, ed. {S}t\'efan van~der {W}alt \& {J}arrod {M}illman, 56 -- 61,
  \dodoi{10.25080/Majora-92bf1922-00a}

\bibitem[{{Wong} {et~al.}(2017){Wong}, {Suyu}, {Auger}, {Bonvin}, {Courbin},
  {Fassnacht}, {Halkola}, {Rusu}, {Sluse}, {Sonnenfeld}, {Treu}, {Collett},
  {Hilbert}, {Koopmans}, {Marshall}, \& {Rumbaugh}}]{Wong2017}
{Wong}, K.~C., {Suyu}, S.~H., {Auger}, M.~W., {et~al.} 2017, \mnras, 465, 4895,
  \dodoi{10.1093/mnras/stw3077}

\bibitem[{{Zhao} {et~al.}(2017){Zhao}, {Song}, \& {Ermon}}]{Zhao2017}
{Zhao}, S., {Song}, J., \& {Ermon}, S. 2017, arXiv e-prints, arXiv:1706.02262.
\newblock \doarXiv{1706.02262}

\bibitem[{{Zhuang} {et~al.}(2019){Zhuang}, {Qi}, {Duan}, {Xi}, {Zhu}, {Zhu},
  {Xiong}, \& {He}}]{Zhuang2019}
{Zhuang}, F., {Qi}, Z., {Duan}, K., {et~al.} 2019, arXiv e-prints,
  arXiv:1911.02685.
\newblock \doarXiv{1911.02685}

\end{thebibliography}
\appendix

\section{Elastic Weight Consolidation}\label{ap:ewc}

Suppose we are given a training set $\mathcal{D}$ and a test task $\mathcal{T}$. The 
posterior of the RIM parameters $\mathcal{\varphi}$ can be rewritten using the Bayes rule as
\begin{equation}
        p(\varphi \mid \mathcal{D},\, \mathcal{T}) = 
        \frac{p(\mathcal{T} \mid \mathcal{D},\, \varphi) p(\varphi \mid \mathcal{D})}
        {p(\mathcal{T} \mid \mathcal{D})}.
\end{equation} 
We suppose that $\varphi$ encode 
information about $\mathcal{D}$, while $\mathcal{T}$ is unseen by $\varphi$. 
It follows that 
$\mathcal{T}$ and $\mathcal{D}$ are conditionally independent when given $\varphi$. 
We do not make the stronger assumption that $\mathcal{D}$ and $\mathcal{T}$ 
are completely independent. In fact, such an assumption 
would contradict the premise of our work that building a 
dataset $\mathcal{D}$ can inform a machine (RIM) about
task $\mathcal{T}$ --- or that, more broadly, $\mathcal{D}$ 
contains information about $\mathcal{T}$.

We rewrite the marginal $p(\mathcal{T} \mid \mathcal{D})$ using the Bayes rule
in order to extract $p(\mathcal{D} \mid \mathcal{T})$, 
the sampling distribution used to compute the Fisher diagonal elements
\begin{equation}
        p(\varphi \mid \mathcal{D},\, \mathcal{T}) = 
\frac{p(\mathcal{T} \mid \varphi) p(\varphi \mid \mathcal{D})}
        {p(\mathcal{D} \mid \mathcal{T})}
        \frac{p(\mathcal{D})}{p(\mathcal{T})}.
\end{equation} 
The log-likelihood $\log p(\mathcal{T} \mid \varphi)$ is equivalent to 
the negative of the loss function for the particular task at hand.
In this work, we assign a uniform probability density to $p(\mathcal{T})$ and $p(\mathcal{D})$ 
in order to ignore them.

We now turn to the prior $p(\varphi \mid \mathcal{D})$, which 
appears as a conditional relative to 
the training dataset. 
We use the Laplace approximation around the maximum $\varphi^{\star}_{\mathcal{D}}$ 
to evaluate the prior,
where $\varphi^{\star}_{\mathcal{D}}$ 
are the trained parameters of the RIM that minimize the empirical risk (equation \eqref{eq:Cost}). 
The Taylor expansion of the prior around this maximum yields
\begin{equation}\label{app:prior}
        \log p(\varphi \mid \mathcal{D}) \approx \log p(\varphi^{\star}_{\mathcal{D}} \mid \mathcal{D}) 
        + \frac{1}{2} (\varphi - \varphi^{\star}_{\mathcal{D}})^{T} 
        \underbrace{
        \bigg(
                \frac{\partial^2 \log p(\varphi \mid \mathcal{D})}{\partial^2 \varphi}\bigg|_{\varphi^{\star}_{\mathcal{D}}}
        \bigg)
}_{\displaystyle \mathbf{H}(\varphi^{\star}_{\mathcal{D}})}
        (\varphi - \varphi^{\star}_{\mathcal{D}}).
\end{equation} 
Since $\varphi^{\star}_{\mathcal{D}}$ is an extremum of the prior, the linear term vanishes. 
The empirical estimate of the negative hessian matrix is the observed Fisher information 
matrix which can be written as
\begin{equation}\label{app:fisher}
        \mathcal{I}(\varphi^{\star}_{\mathcal{D}}) = 
        -\EX_{\mathcal{D} \mid \mathcal{T}} [\mathbf{H}(\varphi^{\star}_{\mathcal{D}})] = 
        \EX_{\mathcal{D}\mid \mathcal{T}}
        \Bigg[
                \Bigg(
                \bigg( 
                        \frac{\partial \log p(\varphi \mid \mathcal{D})}{\partial \varphi}
                \bigg) 
                \bigg( 
                        \frac{\partial \log p(\varphi \mid \mathcal{D})}{\partial \varphi}
                \bigg)^{T}
        \Bigg)
\Bigg|_{\varphi^{\star}_{\mathcal{D}}}\Bigg].
\end{equation} 
The expectation is taken over the sample space $p(\mathcal{D} \mid \mathcal{T})$ since 
the network parameters are held fixed during sampling.
In order to compute the Fisher score, 
we apply the Bayes rule to the prior to extract a loss function,
which we take to be 
proportional to the training loss (equation \eqref{eq:Loss}) and the $\chi^2$:
\begin{equation}\label{eq:LossFisher}
        \log p\big(\varphi \mid (\mathbf{x}, \mathbf{y}) = \mathcal{D}\big) \propto -\mathcal{L}_{\varphi}(\mathbf{x}, \mathbf{y}) + \frac{1}{T}\sum_{t=1}^{T}\log p(\mathbf{y} \mid \mathbf{\hat{x}}^{(t)}) - \frac{\ell_2}{2}\lVert \varphi \rVert^2_2
\end{equation} 
We find in practice the $\ell_2$ term has little effect on the 
Fisher diagonal and our results. Thus, we set $\ell_2 = 0$.

Since the full Fisher matrix is intractable for a neural network, we approximate the 
quadratic term of the prior with the diagonal of the Fisher matrix following \citet{Kirkpatrick2016}. 
For an optimisation problem, the first term of \eqref{app:prior} is constant. Thus,
the posterior becomes proportional to
\begin{equation}
        \log p(\varphi \mid \mathcal{D}, \mathcal{T}) \propto 
        \log p(\mathcal{T} \mid \varphi ) - 
         \frac{\lambda}{2} 
        \sum_{j}\mathrm{diag}(\mathcal{I}(\varphi^{\star}_{\mathcal{D}}))_{j}(\varphi_j - [\varphi^{\star}_{\mathcal{D}}]_j)^2.
\end{equation} 
The Lagrange multiplier $\lambda$ is introduced to tune our uncertainty about the network parameters 
during fine-tuning.

\section{VAE Architecture and optimisation}

For the following architectures, we employ the notion of \textit{level} 
to mean layers in the encoder and the decoder with the same resolution. 
In each level, we place a block of convolutional layers 
before downsampling (encoder) or after upsampling (decoder). These operations 
are done with strided convolutions like in the U-net architecture of the RIM.

\begin{table}[H]
\begin{minipage}{.5\linewidth} 
        \centering
        \caption{Hyperparameters for the background source VAE.}
        \label{tab:Source VAE}
        \begin{tabular}{cc}
                Parameter & Value \\\hline\hline
                Input preprocessing & $\bbone$ \\
                                    & \\

                \textit{Architecture} & \\
                Levels (encoder and decoder) & 3 \\
                Convolutional layer per level & 2 \\
                Latent space dimension & 32\\
                Hidden Activations & Leaky ReLU \\
                Output Activation & Sigmoid \\
                Filters (first level) & 16 \\
                Filters scaling factor (per level) & 2 \\
                Number of parameters & $3\,567\,361$\\

                           & \\
                \textit{Optimization} & \\
                Optimizer & Adam \\
                Initial learning rate & $10^{-4}$ \\
                Learning rate schedule & Exponential Decay \\
                Decay rate & 0.5 \\
                Decay steps & $30\,000$ \\
                Number of steps & $500\,000$ \\
                $\beta_{\mathrm{max}}$ & 0.1 \\
                Batch size & 20\\
                \hline
        \end{tabular}
\end{minipage}
\begin{minipage}{.5\linewidth}
        \caption{Hyperparameters for the convergence VAE.}
        \label{tab:Kappa VAE}
        \begin{tabular}{cc}
                Parameter & Value \\\hline\hline
                Input preprocessing & $\log_{10}$ \\
                              & \\

                \textit{Architecture} & \\
                Levels (encoder and decoder) & 4 \\
                Convolutional layer per level & 1 \\
                Latent space dimension & 16\\
                Hidden Activations & Leaky ReLU \\
                Output Activation & $\bbone$ \\
                Filters (first level) & 16 \\
                Filters scaling factor (per level) & 2 \\
                Number of parameters & $1\,980\,033$\\

                           & \\
                \textit{Optimization} & \\
                Optimizer & Adam\\
                Initial learning rate & $10^{-4}$ \\
                Learning rate schedule & Exponential Decay \\
                Decay rate & 0.7 \\
                Decay steps & $20\,000$ \\
                Number of steps & $155\,000$ \\
                $\beta_{\mathrm{max}}$ & 0.2 \\
                Batch size & 32\\
                \hline
        \end{tabular}
\end{minipage}
\end{table}

\section{RIM architecture and optimisation}\label{ap:rim training and opt}

\begin{figure}[H]
        \centering
        \includegraphics[width=\linewidth]{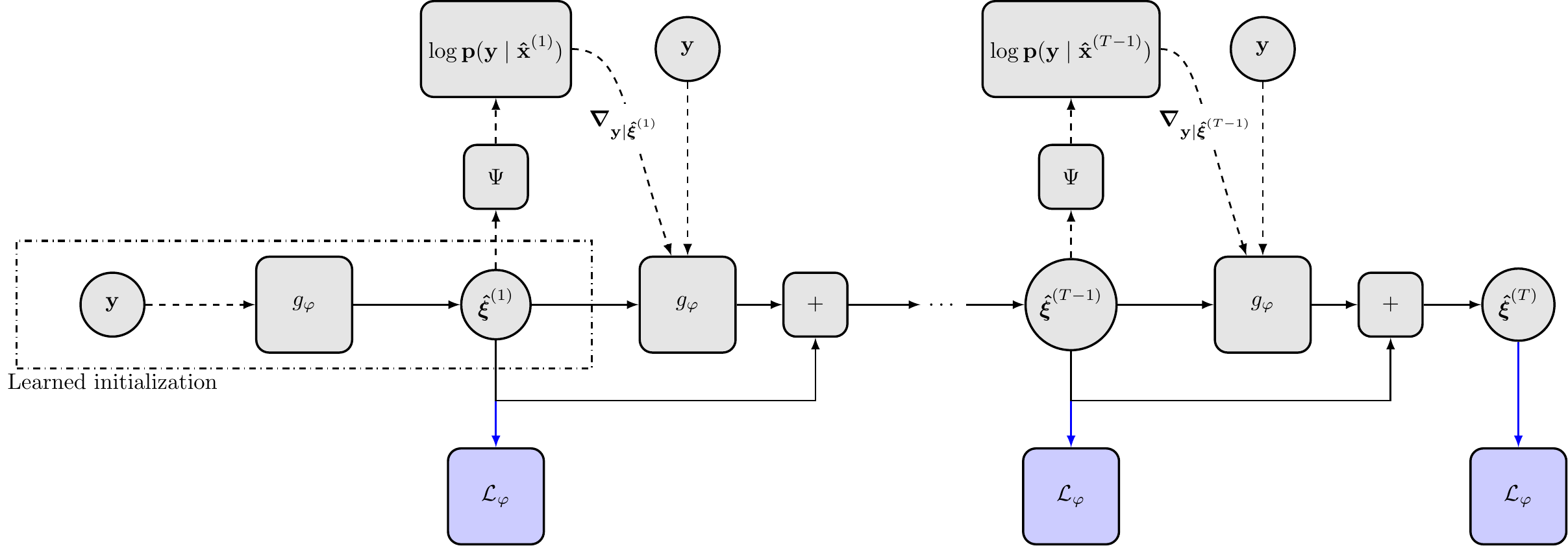}
        \caption{Unrolled computational graph of the RIM. Operations along solid arrows are being 
        recorded for BPTT, while operations along dashed arrows are not. The blue arrows are only 
        used for optimisation during training. During fine-tuning or testing, the loss is computed only 
        as an oracle metric to validate that our methods can recover the ground truth.}
        \label{fig:unrolled graph}
\end{figure}

The notion of a link function $\Psi: \Xi \rightarrow \mathcal{X}$, 
introduced by \citet{Putzky2017}, is an invertible transformation 
between the network prediction space $\boldsymbol{\xi} \in \Xi$ 
and the forward modelling space $\mathbf{x} \in \mathcal{X}$.
This is a different notion from preprocessing, discussed in section \ref{sec:data}, 
because this transformation is applied inside the recurrent relation \ref{eq:RIM} 
as opposed to before training. In the case where the forward model has some restricted 
support or it is found that some transformation helps the training, then 
the link function chosen must be implemented as part of the network architecture as 
shown in the unrolled computational graph in Figure \ref{fig:unrolled graph}.
Also, the loss $\mathcal{L}_\varphi$ must be computed in the $\Xi$ space in order 
to avoid gradient vanishing problems when $\Psi$ is a non-linear mapping, which 
happens if the non-linear link function is applied in an 
operation recorded for backpropagation through time (BPTT). For the convergence, we use an exponential link function with base $10$: 
$\boldsymbol{\hat{\kappa}} = \Psi(\boldsymbol{\xi}) = 10^{\boldsymbol{\xi}}$. 
This $\Psi$ encodes the non-negativity of the convergence. Furthermore, 
it is a power transformation that leaves the linked 
pixel values $\boldsymbol{\xi}_i$ normally distributed, thus improving the 
learning through the non-linearities in the neural network.

The pixel weights $\mathbf{w}_i$ in the loss function \eqref{eq:Loss}
are chosen to encode the fact that the pixels with critical mass density ($\boldsymbol{\kappa}_i > 1$) 
have a stronger effect on the lensing configuration than other pixels. 
We find in practice that the weights 
\begin{equation}\label{eq:convergence weights} 
        \mathbf{w}_i = \frac{\sqrt{\boldsymbol{\kappa}_i}}{ \sum_i \boldsymbol{\kappa}_i}, 
\end{equation} 
encode this knowledge in the loss function and improve both the empirical 
risk and the goodness of fit of the baseline model on early test runs.

For the source, we found that we do not need a link function 
--- the identity is generally better compared to other link functions we tried like sigmoid and power transforms --- and we found that the pixel weights can be taken to 
be uniform, i.e.\ $\mathbf{w}_i = \frac{1}{M}$.

\begin{table}[H]
        \centering
        \caption{Hyperparameters for the RIM.}
        \label{tab:baseline hparams}
        \begin{tabular}{cc}
                Parameter & Value \\\hline\hline
                Source link function & $\bbone$ \\
                $\kappa$ link function & $10^{\boldsymbol{\xi}}$ \\
                                       & \\
                \textit{Architecture} & Figure \ref{fig:unet} \\
                Recurrent steps ($T$) & 8 \\
                Number of parameters & $348\,546\,818$ \\
                                      & \\
                \textit{First Stage Optimisation} & \\
                Optimizer & Adamax \\
                Initial learning rate & $10^{-4}$\\
                Learning rate schedule & Exponential Decay \\
                Decay rate & 0.95 \\
                Decay steps & $100\,000$\\
                Number of steps & $610\,000$\\
                Batch size & 1 \\
                           & \\
                \textit{Second Stage Optimisation} & \\
                Optimizer & Adamax \\
                Initial learning rate & $6\times 10^{-5}$\\
                Learning rate schedule & Exponential Decay \\
                Decay rate & 0.9 \\
                Decay steps & $100\,000$\\
                Number of steps & $870\,000$\\
                Batch size & 1 \\
                
                \hline
        \end{tabular}
\end{table}

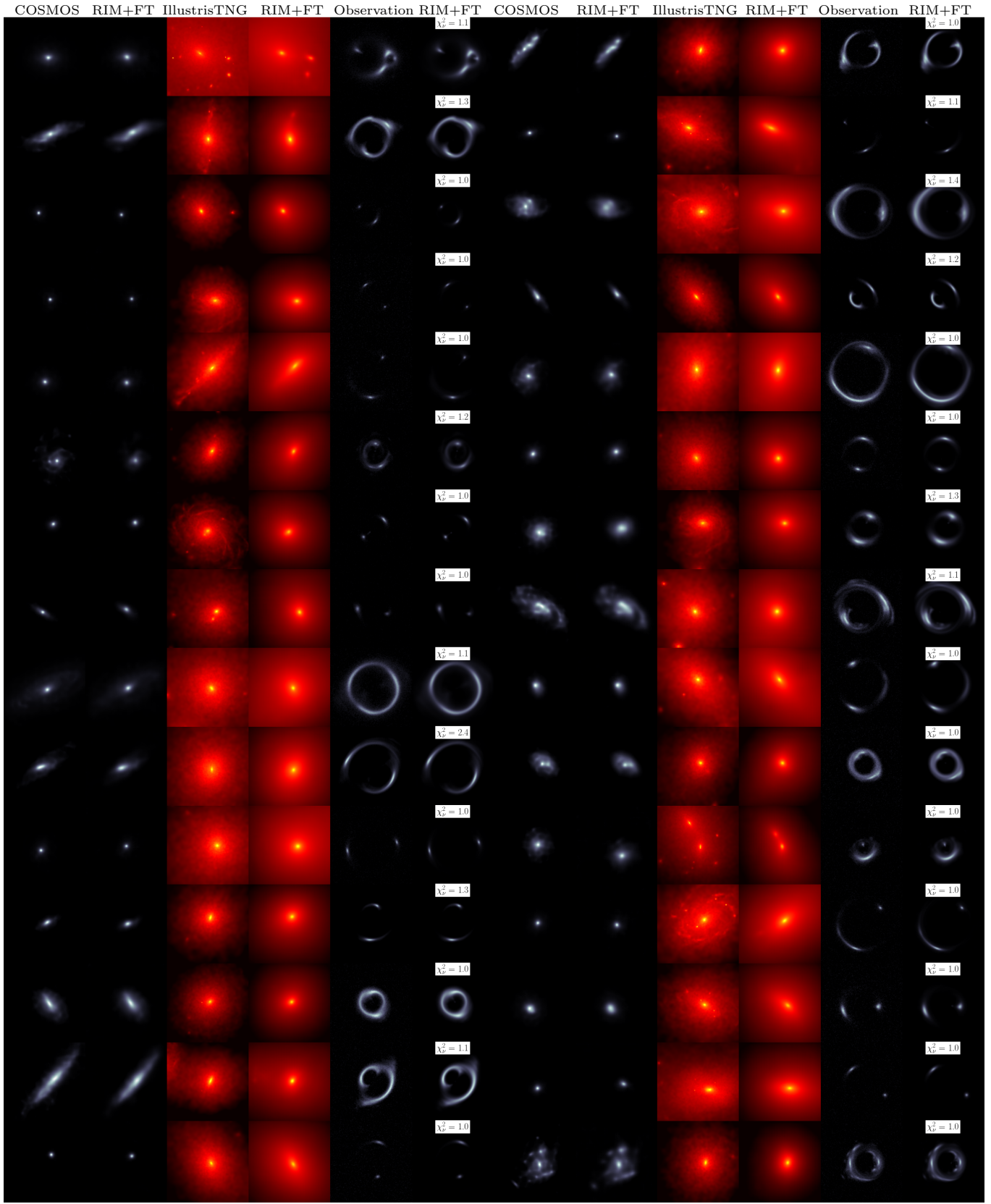
\begin{figure}[H]
        \centering
        \begin{tikzpicture}
                \tikzstyle{every node}=[font=\scriptsize]
                \node at (0, 0) {\includegraphics[width=\linewidth]{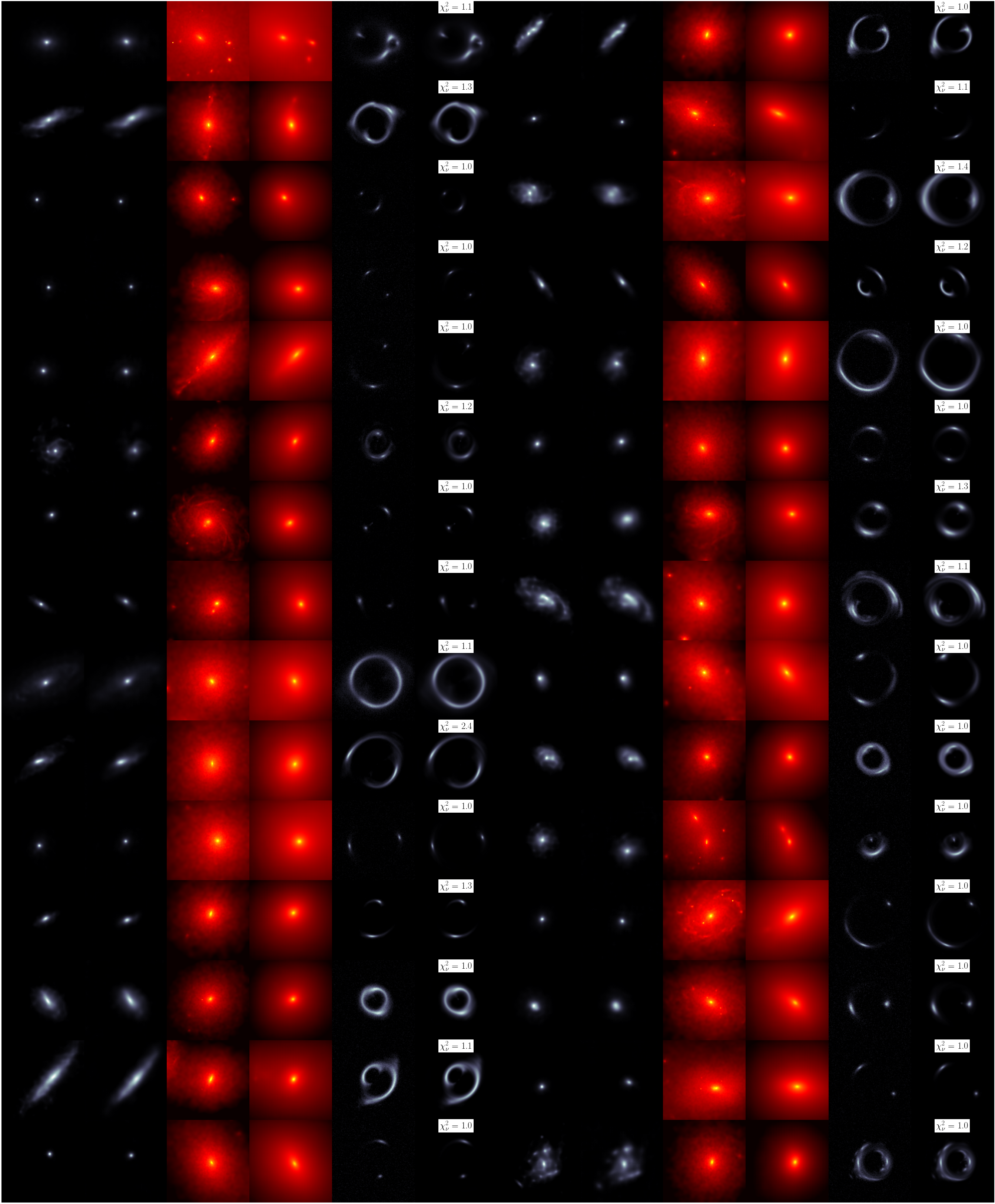}};
                \node at (-8.2, 11) {COSMOS\strut};
                \node at (-6.8, 11) {RIM+FT\strut};
                \node at (-5.3, 11) {IllustrisTNG\strut};
                \node at (-3.7, 11) {RIM+FT\strut};
                \node at (-2.2, 11) {Observation\strut};
                \node at (-0.8, 11) {RIM+FT\strut};
                \node at (0.6, 11) {COSMOS\strut};
                \node at (2.1, 11) {RIM+FT\strut};
                \node at (3.7, 11) {IllustrisTNG\strut};
                \node at (5.2, 11) {RIM+FT\strut};
                \node at (6.7, 11) {Observation\strut};
                \node at (8.2, 11) {RIM+FT\strut};
        \end{tikzpicture}
        \caption{
                30 reconstructions taken at random from the test set of 3000 examples simulated from COSMOS 
                and IllustrisTNG data at high SNR.
                The colorscale are the same as in Figure \ref{fig:main result}.}
        \label{fig:random sample}
\end{figure}


\end{document}